# Darknet Traffic Analysis: A Systematic Literature Review

Javeriah Saleem[1*], Rafiqul Islam[1], Zahidul Islam[1]

[1]School of Computing and Mathematics, Charles Sturt University, Albury, NSW, Australia

*Corresponding author: Javeriah Saleem (e-mail: javeriahsaleem1@gmail.com)

**ABSTRACT** The primary objective of an anonymity tool is to protect the anonymity of its users through the implementation of strong encryption and obfuscation techniques. As a result, it becomes very difficult to monitor and identify users' activities on these networks. Moreover, such systems have strong defensive mechanisms to protect users against potential risks, including the extraction of traffic characteristics and website fingerprinting. However, the strong anonymity feature also functions as a refuge for those involved in illicit activities who aim to avoid being traced on the network. As a result, a substantial body of research has been undertaken to examine and classify encrypted traffic using machine-learning techniques. This paper presents a comprehensive examination of the existing approaches utilized for the categorization of anonymous traffic as well as encrypted network traffic inside the darknet. Also, this paper presents a comprehensive analysis of methods of darknet traffic using ML (machine learning) techniques to monitor and identify the traffic attacks inside the darknet.

## I. INTRODUCTION

The term "dark web" refers to a stratum below the deep web within the internet protocol stack. This particular layer remains inaccessible to conventional search engines, such as Google, YouTube, Yahoo, Bing, Baidu, and Yandex, as they cannot index its content. The limitation of conventional search engines to conducting searches exclusively on the surface web has resulted in the dark web becoming a sanctuary for orchestrating and perpetuating various cybersecurity threats. The surface web constitutes approximately 5% of the total web and can be readily accessed through conventional search engines. However, the dark web, comprising the remaining portion, necessitates the utilization of specialized software, such as The Onion Router (Tor), Freenet, Jondonym, whonix, Riffle, and Invisible Internet Project (I2P) for accessibility.

The dark web has garnered substantial attention in both national and international media due to its reputation as a vast black market where various cybersecurity threats are orchestrated. The aforementioned items found within the mentioned context are of questionable authenticity, including counterfeit currency, forged passports, and falsified identity documents. The platform in question has evolved into a notable hub for the illicit trade of lethal armaments, contraband narcotics, unauthorized disclosure of sensitive data, and the dissemination of explicit material involving minors. According to a recent report published by CloudFlare, most requests originating from the Tor browsing network, specifically 94%, exhibit a notable inclination toward cybersecurity threats.

The darknet refers to a segment of the internet that encompasses unused address space and is intentionally designed to operate independently from the rest of the global computer network. It is characterized by its deliberate isolation from external connections and is not intended for conventional interaction with other computers worldwide. The nomenclature "dark" has been attributed to this particular entity due to its inherent characteristic of anonymity, as well as its function as a virtual marketplace facilitated by the utilization of cryptocurrency. Communication originating from the dark space is regarded with skepticism due to its inherent passive listening characteristic, whereby it solely accepts incoming packets while lacking support for outgoing packets.

In the context of the darknet, where legitimate hosts are not present, any incoming traffic is generally regarded as undesired and is typically categorized as a probe, backscatter, or misconfiguration. Darknets, alternatively referred to as network telescopes, sinkholes, or blackholes, are recognized as distinct entities within computer networks. The variation in traffic across different darknets is notably influenced by the magnitude of the IP range designated for surveillance purposes. The size of the darknet exhibits considerable variability, ranging from a solitary host to encompassing the expanse of the largest available IP address space. The advent of sophisticated privacy tools to access the darknet has presented a notable obstacle for law enforcement agencies (LEAs) in their ability to efficiently detect and bring to justice individuals involved in cybercriminal activities. The capacity of these individuals to effectively obscure their identities and actions on the internet has presented a significant challenge for LEAs in their endeavors to address and mitigate cybercrime.

Two main classifications for anonymity systems in the darknet are high-latency systems and low-latency systems. Systems with high latency, such as The Second-Generation Onion Router [8] and the Mixmaster protocol [18], offer superior protection against attacks that exploit packet timing. These systems utilize various techniques, such as mixing, reordering, and patching, to protect against traffic analysis attacks that exploit packet timings and delays. The adoption of high-latency anonymity solutions is limited due to the additional delays they introduce in the transmission of data.

In contrast, low-latency systems abstain from employing methods that introduce communication delays, making them well-suited for online surfing protocols like HTTP and interactive protocols such as SSH. This category of anonymity systems encompasses Tor (The Onion Routing) [1], Java Anon Proxy (JAP) [20], and Invisible Internet Protocol (I2P) [21]. Other than these, anonymity can be achieved by using proxy, VPN, and DNS. The utilization of anonymity solutions by users extends to both lawful and illicit actions. The analysis of darknet traffic plays a crucial role in the proactive monitoring of malware, enabling researchers to detect and mitigate potential threats before they can cause significant harm. Additionally, it aids in identifying and investigating malicious activities that have already occurred, allowing for a more comprehensive understanding of the outbreak and facilitating appropriate countermeasures. The impetus behind this research can be attributed to several key motivations.

The systematic review of scientific literature is significant in identifying research questions and justifying future research in a specific study area. The systematic literature review (SLR) is a research methodology that seeks to identify and analyze relevant works within a specific study area. It employs a systematic approach, adhering to predetermined research steps and processes. The primary objective of an SLR is to comprehensively gather and evaluate existing literature to synthesize and summarize the current state of knowledge on a particular topic. Despite the existence of studies that have examined the intricacies of darknet traffic, a comprehensive and systematic literature review pertaining to the monitoring, detection, and attacks within the Dark Web specifically in relation to traffic remains inadequately explored. This dearth of research has served as the impetus for our endeavor to present this survey, aiming to address this knowledge gap. The primary objective of this study is to investigate the various techniques, models, and methods that are currently being developed and utilized for identifying and categorizing darknet traffic, as well as detecting and classifying malicious activities and attacks through the analysis of network traffic. To fulfill the objectives of our study, a meticulous and methodical approach was employed to identify and analyze a total of 66 scholarly articles deemed pertinent to our research focus. The contributions of this paper can be summarized as follows:

- It provides a comprehensive review of the current literature (2017–2023) using a SLR.
- The presented review involved a detailed examination of the taxonomy of darknet traffic analysis.
- This study presents a comprehensive analysis of the detection methods employed for identifying darknet traffic attacks through a SLR.
- It has identified various challenges and complexities associated with darknet traffic analysis.

The remainder of this paper is organized as follows:

Section II presents the research methodology using a SLR. Section III presents demographic information from the literature. Section IV presents the architectural designs employed in the chosen articles. Section V presents the findings and analysis, while Section VI presents the challenges in darknet traffic analysis. Section VII discusses research gaps. Section VIII concludes our study, encompassing a discussion of our findings and potential avenues for further research in this field.

## II. RESEARCH METHODOLOGY

The systematic literature review (SLR) methodology that was employed to carry out this review is described in this section. We have also considered a few recent studies using the SLR approach. SLR employs systematic procedures to formulate the research topic, conduct the literature search, screen the results, extract the data from the chosen results, and then qualitatively or quantitatively analyze and synthesize the results. Determining the research questions, suitable data sources, search techniques, inclusion and exclusion criteria, data extraction, analysis, and synthesis are all part of the approach.

*A. RESEARCH QUESTIONS*

The internet's core infrastructure and end-users' digital assets have suffered significant damage due to the growth of harmful activity. Monitoring darknets or unused blocks of Internet Protocol addresses is an inexpensive method of keeping tabs on global cybercrime. Careful monitoring and analysis of darknet network data can help address and lessen the impact of criminal activity on the underground web. This highlights the need for more effective monitoring systems and law enforcement.

The primary objective of this work is to provide an overview of Dark Web traffic analysis, along with detection and intelligence techniques. This study seeks to elucidate the various law enforcement methodologies and technological tools employed in the pursuit of tracing and detecting darknet traffic. This statement posits that the utilization of contemporary technologies, in conjunction with law enforcement efforts, presents a prospective trajectory for implementing diverse strategies to mitigate cyber threats. The primary aim of this study is to examine the utilization of machine learning methodologies for the classification of encrypted darknet data, as well as to review recent studies on the classification of encrypted traffic on Clearnet networks. In light of the findings from the research, we will engage in a full discussion and comparison of various machine-learning techniques and their respective operations. The research aims to answer the following research questions:

- RQ1 What are the main themes of the publications in the darknet traffic analysis domain?
- RQ2 What is the process of darknet traffic detection and classification?
    - RQ2.1 What datasets are used to perform the darknet traffic classification?
    - RQ2.2 How does feature selection impact the traffic analysis?
    - RQ2.3 What are common ML algorithms used in the darknet traffic analysis domain?
    - RQ2.4 What metrics are used to measure classification accuracy within the traffic classification domain?
- RQ3 What is the taxonomy of darknet traffic analysis?
    - RQ3.1 How can malicious activities be detected through darknet traffic analysis?
    - RQ3.2 What countermeasures can be taken to avoid the deanonymization of darknet traffic?
- RQ4 What are the challenges in the darknet traffic analysis domain?
- RQ5 What are the research gaps and future research options?

*B. SEARCH STRATEGY AND SELECTION*

The guidelines for performing a systematic literature review have been followed in accordance with the search strategy, as detailed in the subsequent section. To get the essential data for the review papers, an extensive electronic search was performed on respected academic databases, such as IEEE Xplore, Scopus, ACM Digital Library, and Google Scholar. The terminology utilized in the research queries of our study has been integrated into the corresponding area. Boolean search procedures, namely employing the logical operators "AND" and "OR," have been implemented for specific phrases. Table I presents the search terms utilized for retrieving publications relevant to our investigation. It is noteworthy to acknowledge that diverse search terms have been utilized to retrieve relevant scholarly publications. Moreover, a thorough analysis of the references included in the pertinent publications was undertaken to ascertain other academic sources. The procedure for matching strings within search terms in digital libraries is predicated upon scrutinising the title, abstract, and keywords linked to the publications. The procedure of filtering and screening was carried out to find the most relevant papers, following the established criteria for inclusion and exclusion. The survey's criteria for inclusion and exclusion are described in Table II and Table III, respectively. Following the implementation of the aforementioned screening methods, which entailed the utilization of specific inclusion and exclusion criteria, 66 papers were selected for inclusion in the present review study. Figure 1 shows the literature selection from the database libraries.

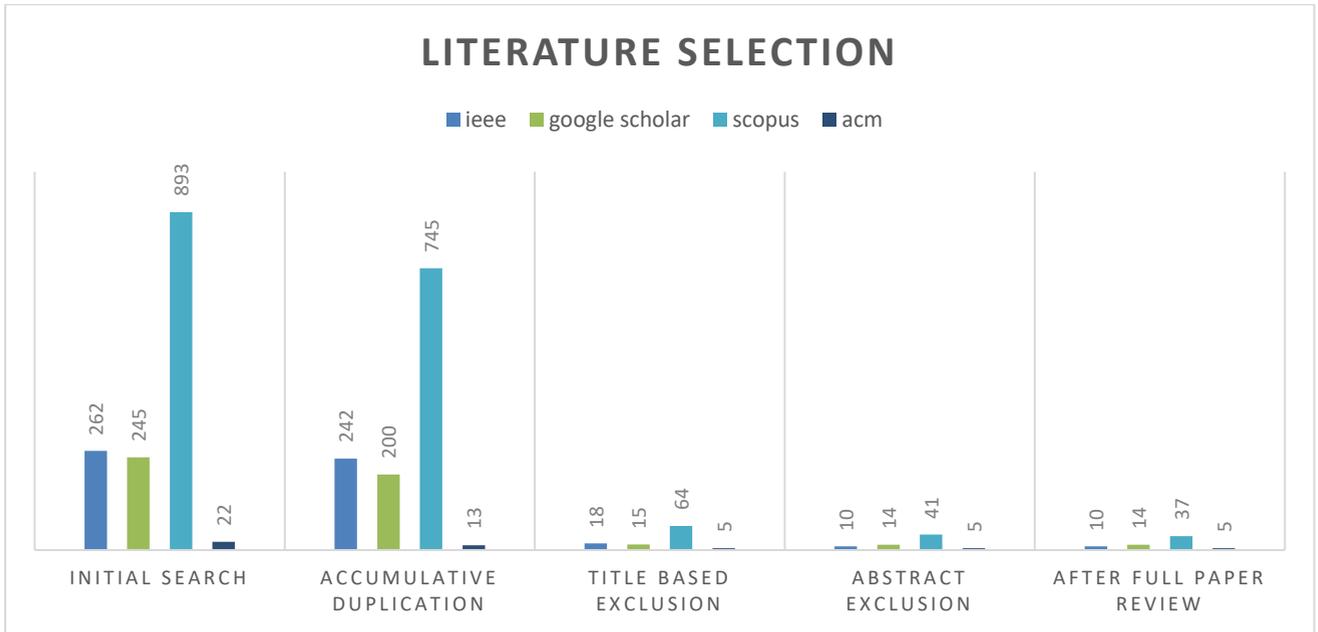

Figure 1 Literature Selection Process

**Automatic search**: An automated search was conducted using the search criteria stated in the four database libraries mentioned earlier, resulting in a total of 1422 papers being obtained.

**Removal of duplicate papers**: In this particular instance, the removal of duplicate papers was undertaken due to the presence of certain database index documents that are accessible through alternative databases. The total count of articles was reduced to 1200 subsequent to the elimination of duplicate entries.

**Title-based selection:** The strategy employed for expeditious article selection used a title-based approach. We have excluded papers from our systematic literature review that do not contain the phrases "traffic analysis," "traffic attacks," "traffic monitoring," or "traffic detection" in their titles, as they are not relevant to our research. The aforementioned action resulted in the cumulative count of papers reaching 102.

**Abstract-based selection:**. The relevance of the 102 papers' abstracts to our systematic literature review was assessed. At this juncture, the abstract articles deemed unnecessary were disregarded, resulting in the selection of 70 papers.

**Full-paper review:** The entire set of 66 papers was comprehensively reviewed, resulting in the selection of 10 publications from IEEE Xplore, 14 from Google Scholar, 37 from Scopus, and the remaining 5 from the ACM Digital Library.

| TABLE I Search Terms For A Literature Selection ||
|---|---|
| S# | Search Term |
| 1 | ("Darknet" AND "traffic analysis") OR ("Darknet" AND "traffic attacks") OR ("Darknet" AND "traffic monitoring") |
| 2 | ("Tor" AND "traffic analysis") OR ("Tor" AND "traffic attacks") OR ("Tor" AND "traffic monitoring") |
| 3 | ("Darkweb" AND "traffic analysis") OR ("Darkweb" AND "traffic attacks") OR ("Darkweb" AND "traffic monitoring") |

| TABLE II Inclusion Criteria For Selection Of Literature ||
|---|---|
| IC# | Inclusion Criteria |
| IC1 | A study that is related to the darknet or Tor network. |
| IC2 | The search term keywords in TABLE I have an AND operator showing both key terms must be present in the search whereas the OR operator means at least one of the key terms should be in the search. |
| IC3 | Study published from 2017–2023. |
| IC4 | The study must be in the English language. |
| IC5 | All journal conference and survey articles are included in this review. |
| IC6 | A study that has traffic analysis, traffic detection or traffic monitoring words in the title of the research work. |
| IC7 | Included abstract and full-text. |

| TABLE III Exclusion Criteria For Literature ||
|---|---|
| EC# | Exclusion Criteria |
| EC1 | The title doesn't have key terms like "Tor traffic analysis" "Tor traffic attacks" "Darknet Or dark web traffic analysis OR attacks OR monitoring" |
| EC2 | Exclude the duplication of articles obtained from different databases |
| EC3 | The abstract is not related to the literature review research area |

| TABLE IV The QARs Of The SLR ||
|---|---|
| QAR# | Research Questions Description |
| QAR1 | Are the objectives of the research questions clearly defined? |
| QAR2 | Is the article taking current and past literature into account? |
| QAR3 | Are the methods used to analyze the results clear? |
| QAR4 | Are the darknet traffic detection techniques clearly defined? |
| QAR5 | Has the study mentioned any traffic attack or its detection technique? |

*C. DATA EXTRACTION AND SYNTHESIS*

This part discusses the technique of data extraction and analysis of the data obtained from the filtered publications to address the research questions in this systematic literature review. The process of extracting data from the filtered articles was conducted using the data extraction form presented in Figure 2. The data extraction process involved the utilization of a Microsoft Excel spreadsheet for recording the acquired data.

To assess the appropriateness of an article in relation to addressing the research topic, the quality attribute rules were utilized. Five QARs were identified, with each QAR having a value of 1. The cumulative score of the article will be determined by the aggregation of marks acquired from the five Quality Assessment Rubrics (QARs). Articles that yielded a result of 3 or above were deemed to have adequately addressed our research questions, while those that did not meet this threshold were omitted from further analysis. The QARs are presented in Table IV. Figure 3 explains the complete literature selection criteria for this SLR.

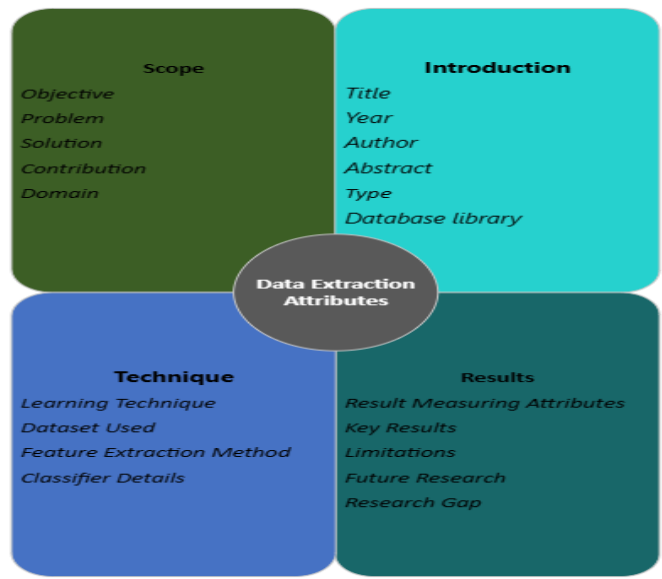

*Figure 2 Data Extraction Attributes*

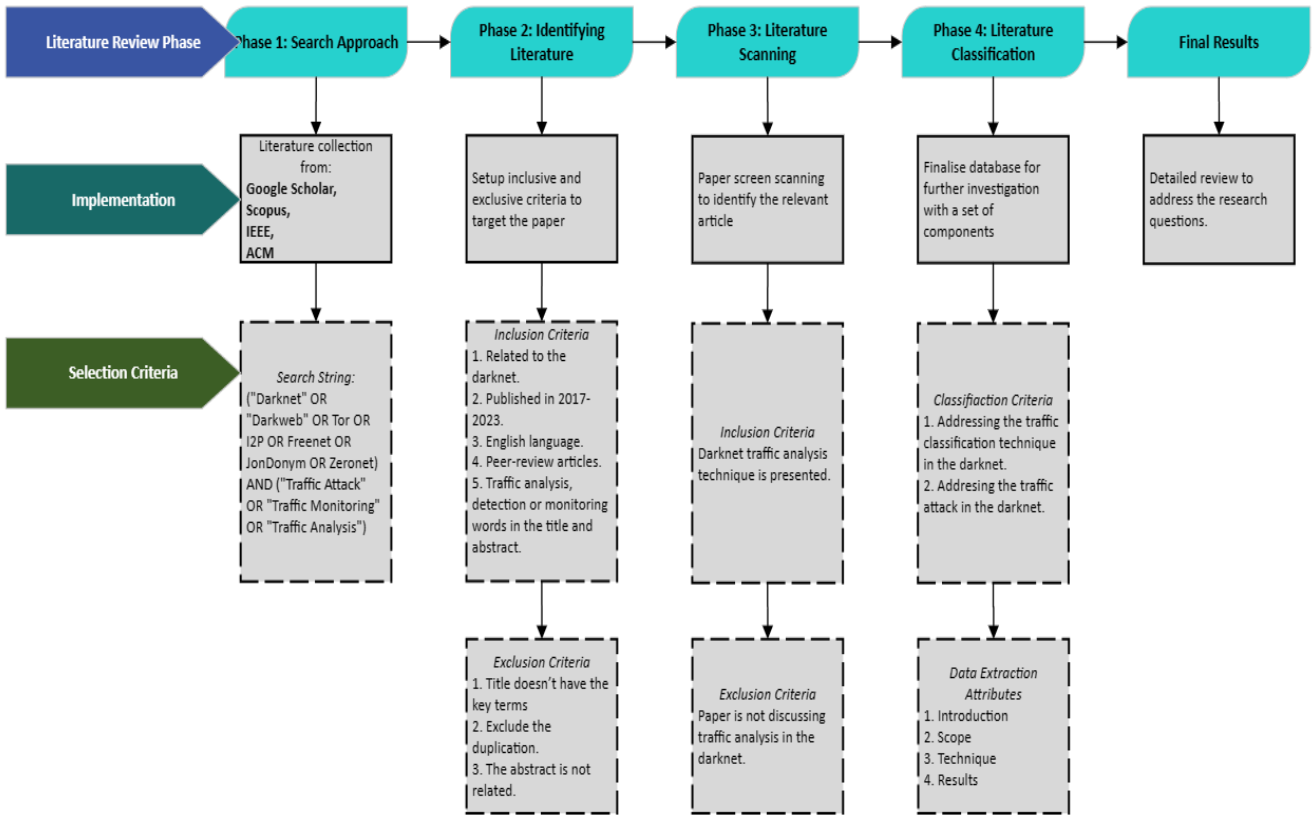

*Figure 3: Literature Selection Criteria*

For data synthesis on the extracted data based on the quality features presented in Table IV, information is extracted based on the attributes outlined in Figure 2. We categorized our literature evaluation according to the research questions, considering the objectives and motives stated in the publications. Subsequently, thematic analysis was employed to extract the underlying themes from the aforementioned investigations. A qualitative data analysis was conducted, utilizing the themes that were derived. The present study provides an overview of the architecture frameworks utilized in the existing literature, organized according to their respective thematic categories. TABLE V gives an elaborated overview of the researchers and their respective areas of work. These studies highlight the detection of anonymous traffic as one of the traffic analysis challenges associated with the darknet. After identifying the generalized models through thematic analysis in Figure 6, we analyzed the proposed models to determine the various tools and techniques employed inside these models. Subsequently, we applied the framework analysis mentioned in Figure 8 to derive generalized methods from the aforementioned analysis. This step functions as the elucidation and exemplification phase in examining the architecture.

This study thoroughly examines the existing body of research about network traffic analysis and inspection, specifically within the framework of the growing prevalence of network traffic encryption. It investigates the current advancements in the field and evaluates the existing literature that presents potential approaches for conducting inspections in situations where network communication is encrypted.

## III. DEMOGRAPHIC INFORMATION

This section presents an analysis of the distribution of the reviewed articles according to the types of publications, publications by year, and publication classification, as determined by the specified search parameters.

### a. CHRONOLOGICAL PERSPECTIVE

According to Figure 1, the literature under consideration was published between 2017 and 2023. Our systematic literature review (SLR) analysis included publications published before September 2023. These papers were thoroughly examined as part of our comprehensive search approach to identify relevant studies. Figure 3 illustrates an upward trend in publications about the Dark Web domain. Our literature analysis shows Scopus accounts for most sources, whereas IEEE published the most research in the area in 2022.

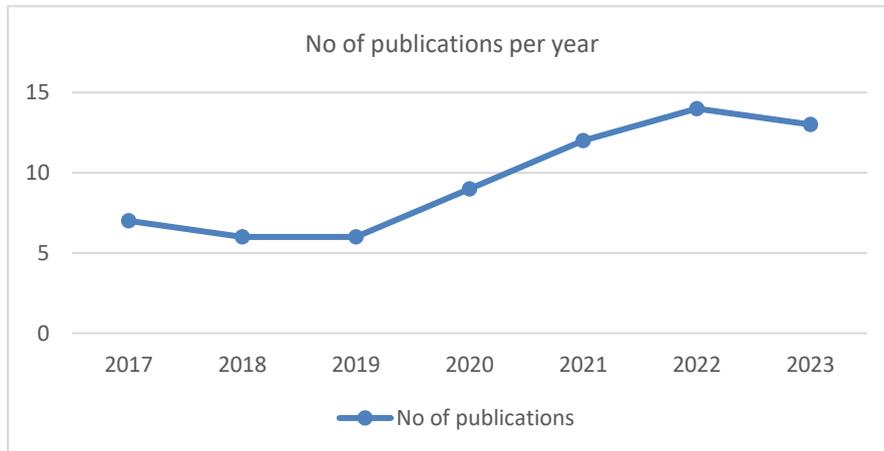

*Figure 4 Chronological View*

**PUBLICATION CRITERIA**

Figure 5 illustrates the distribution of selected review papers based on criteria. The primary research questions (RQs) are adequately addressed in most papers based on the specified criteria. The other criteria are also pertinent to our research questions. The study articles encompass several topics related to Dark Web traffic detection in Tor, I2P, Freenet, JonDonym, Zeronet, and other malicious activities along with the attacks on darknet traffic.

b. **LITERATURE MAPPING**

It is essential to consider the interrelated nature of this research and the fact that these studies share certain information in common. The mapping of all the different pieces of literature and their interconnections is shown in Figure 6. We have conducted a study on a subset of 66 papers with the most citations per year. Based on our investigation, it has been determined that Lashkari possesses the paper with the highest number of citations in this particular field of study. While it is true that certain writers have written works in several fields of study, our assessment of their popularity is primarily based on the dataset he made publicly available.

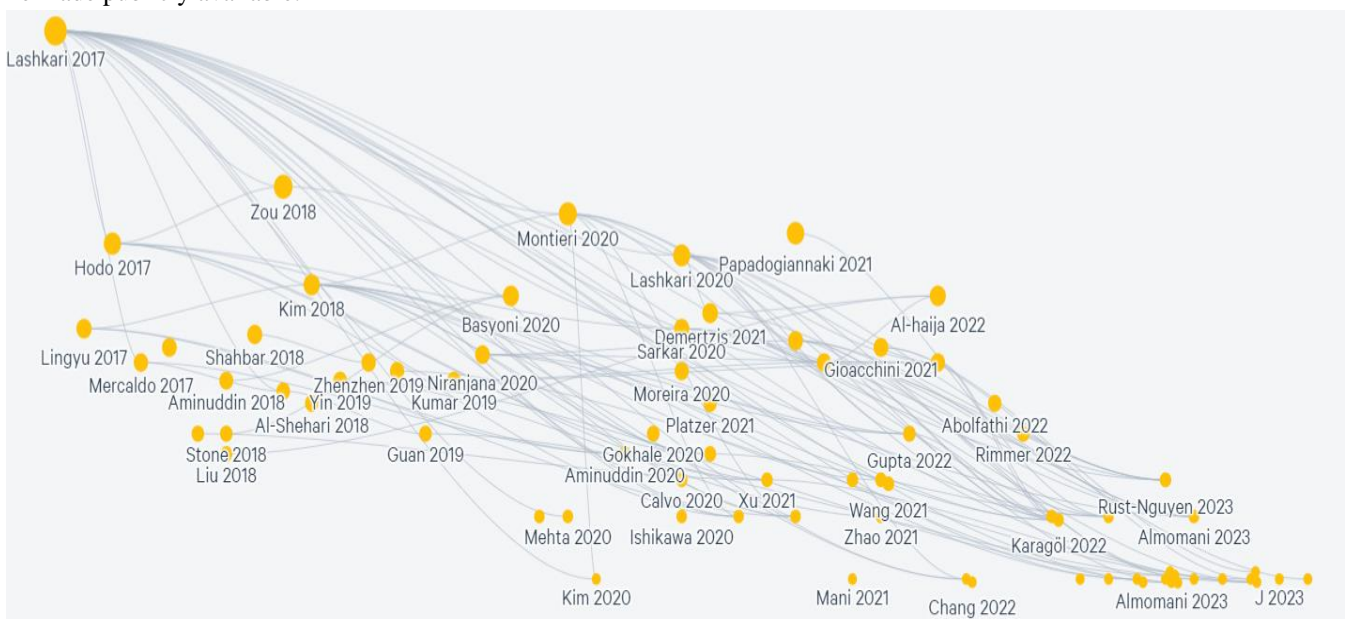

*Figure 5 Literature Mapping*

## IV. ARCHITECTURAL FRAMEWORK (RQ1)

This section describes the architectural frameworks and gives architectural descriptions of the selected articles from the darknet. Figure 6, Figure 7, and Figure 8 depict a comprehensive aerial perspective of the literature. It should be noted that the various components of the designs are not restricted to what is represented in the image. This is something that should be kept in mind. The publications that we examined provided a variety of architectural frameworks, each of which was predicated on a unique traffic analysis. The most in-depth discussion of all of the specifics can be found in Sections V & VI. Figure 6 is the example that will be used to illustrate how the overall structures of the papers should be analyzed. Researchers are concentrating their attention on various topics, including Tor, I2P, Freenet, and others. The data extracted from the attributes mentioned in Figure 2 served as the inspiration for the development of the architectural framework. The standardized procedures that were employed in the models that were applied to the various designs are outlined in Figure 8. This graphic makes it easier to comprehend the important parts of a model, as well as their instances and significance. After analyzing all the information obtained from the architectural framework analysis, we came up with the response for our RQ1. TABLE V presents the main themes and the literature reviewed in the particular category based on the purpose of the theme, whereas Figure 9 presents the percentage classification of the reviewed literature. The framework may be broken down into two distinct categories: the first is for traffic detection methods, and the second is for threat analysis.

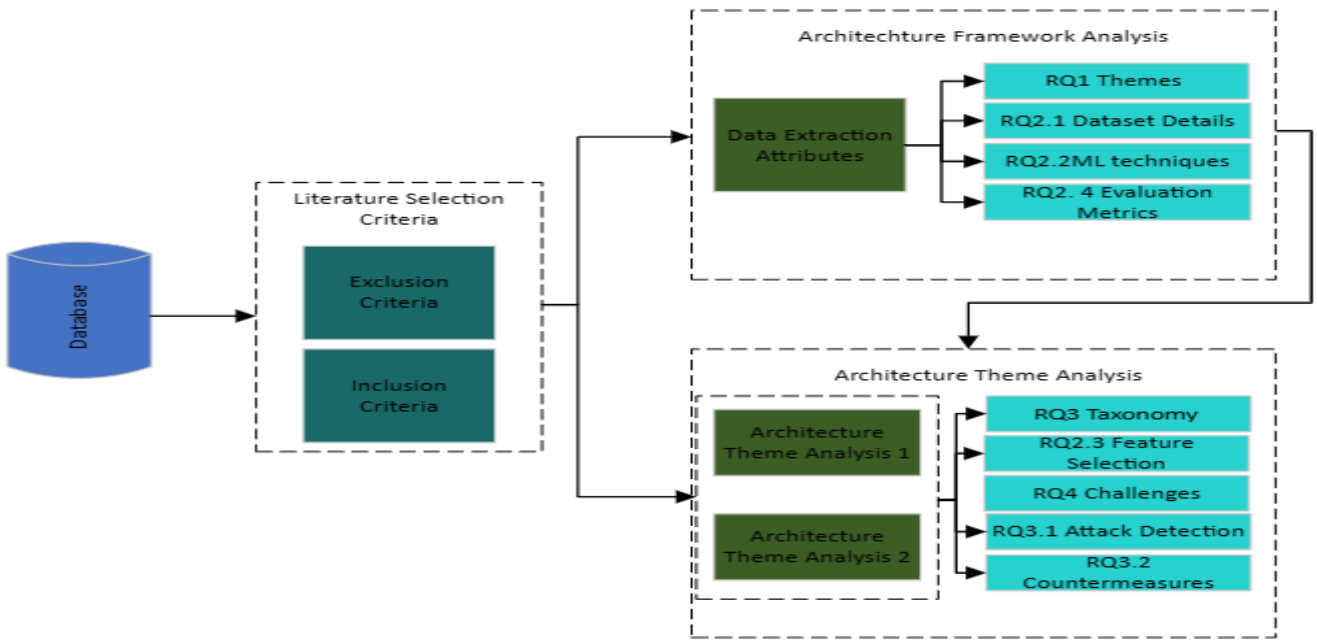

*Figure 6 Data Analysis Process*

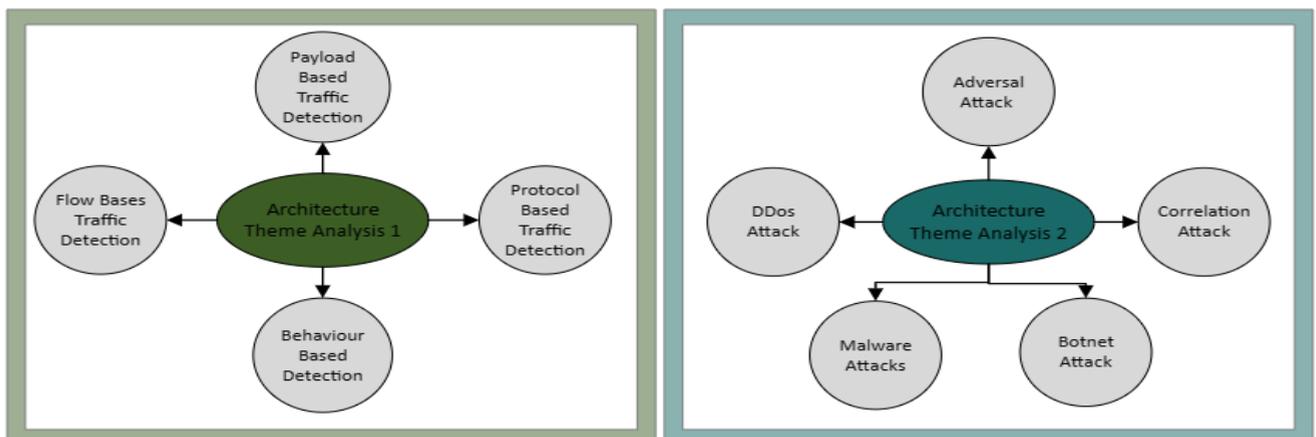

*Figure 7 Architecture Theme Analysis*

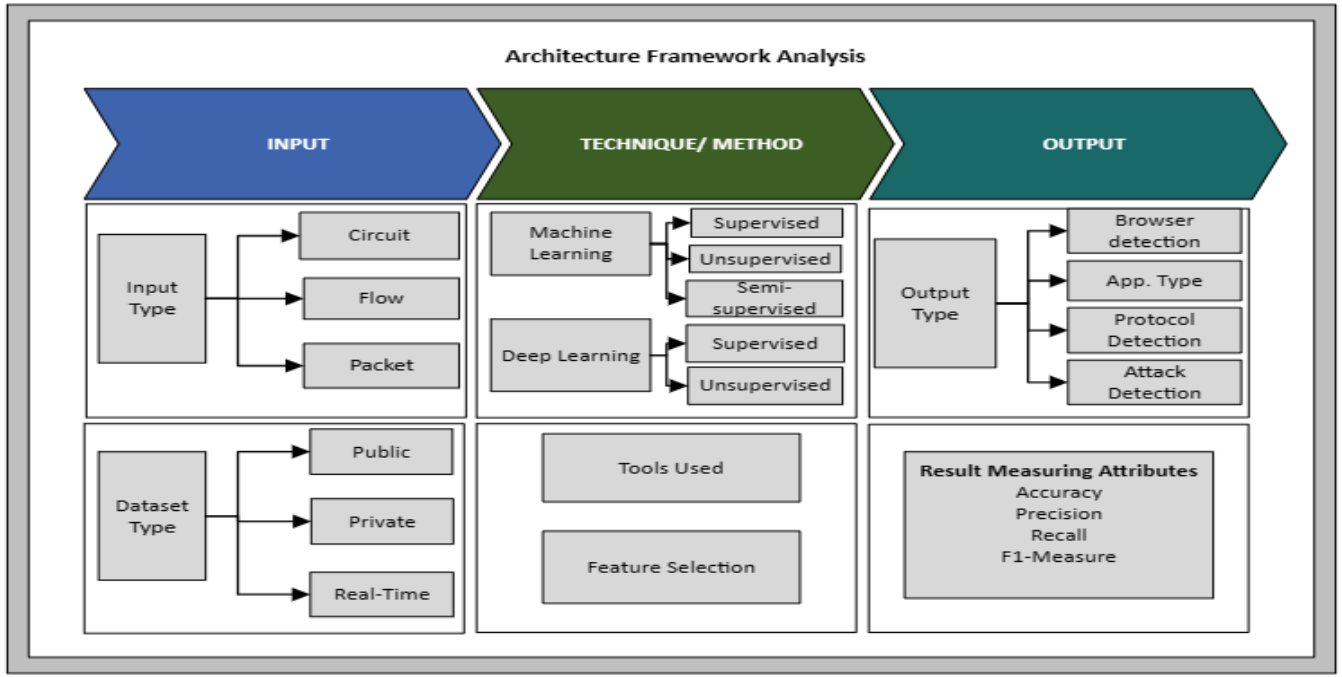

*Figure 8 Architecture Framework Analysis*

| | TABLE V Summary of Main Themes of The Surveyed Literature | | |
|---|---|---|---|
| Analysis Type | Theme | Description | References |
| Browser-based | Obfuscated traffic | Study on obfuscated Tor traffic analysis | [1], [2], [3] |
| | Non-obfuscated traffic | Study on non-obfuscated traffic analysis | [4],[5], [6], [7], [8], [9], [10], [11], [12], [13], [14], [15], [16], [17] |
| | Multiple browser classification | Study classifying the darknet traffic into multiple browsers | [18], [19], [20] |
| | Browser settings | Classification of the darknet browser through its settings | [21], [22], [23], [24], [25] |
| | Padded traffic detection | Classification of traffic after applying the defense mechanism | [6] |
| | Traffic classification under adversarial settings | Classification of Tor traffic under adversarial settings | [1], [26] |
| Application-based | Classification of applications | Classifying darknet traffic and applications used by it, e.g., audio, video, browsing etc. | [27], [28], [29], [30], [31], [22], [32], [33], [34] |
| | Classification of fine-grained applications | Further classification of darknet applications into software used by the applications, e.g., Facebook, Twitter, BitTorrent etc. | [35], [36], [17] |
| Protocol-based | Protocol-based traffic analysis | Classified the protocols used in the darknet | [37], [38] |
| Behavior-based | Attack prediction/detection | Attack prediction or detection through analyzing traffic patterns | [39], [40], [41], [42], [43], [44], [45], |
| | Traffic attacks | Studies on new proposed traffic attacks | [46], [37], [47] |
| | Relay detection | Deanonymizing Tor traffic through relay detection | [48], [49] |

| | Counterattack | Counter attack techniques in the dark web | [50], [51], [52] |
|---|---|---|---|
| | Real-time attack detection | Applying attack detection techniques on real-time data | [53], [54], [55] |
| Data balancing | Impact of data balancing | Studies focusing on data balancing techniques and their impact on darknet traffic classification | [56], [57] |
| Feature selection | Feature selection algorithm | Studies focusing on the feature selection for the classification purpose | [58], [57], [12], [59] |

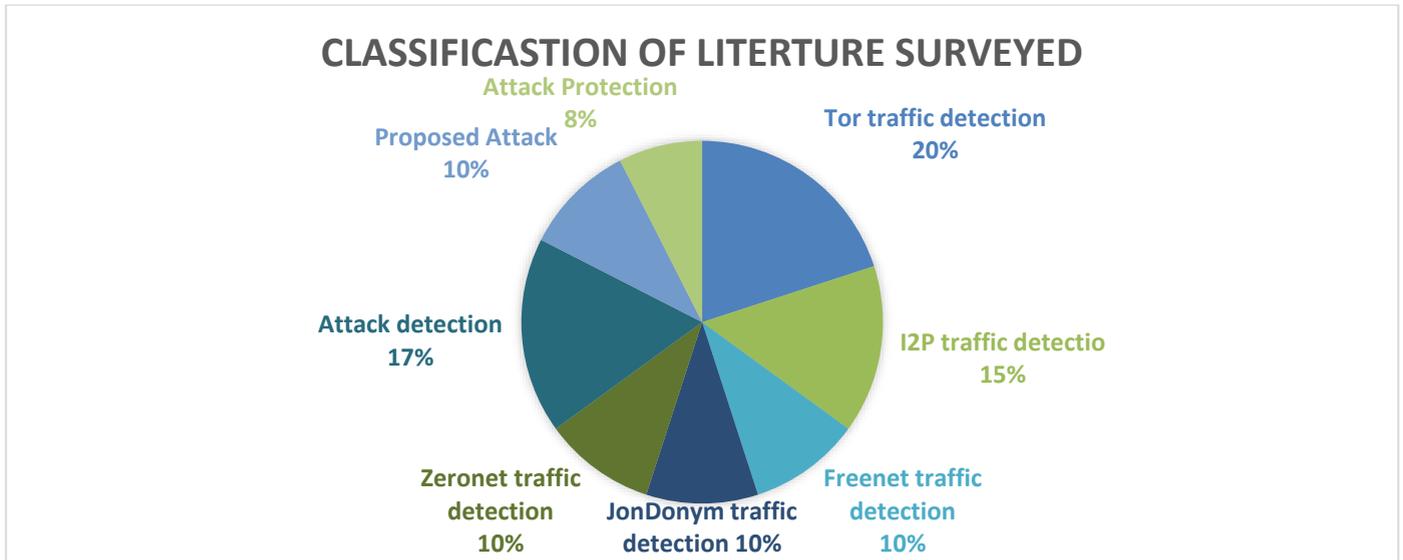

Figure 9 Percentage classification of the literature reviewed

## V. INSIGHTS FROM THE DARKNET TRAFFIC ANALYSIS (RQ2)

This section discusses RQ2 and its subsequent research inquiries and evaluates the outcomes from systematically examining the scholarly literature. Section A discusses the major darknet browsers used in the research. The investigation will address the major research question (RQ2) in the framework of subsection B. The discussion of the response to the sub-parts of RQ2 is presented in subsections C, D, E, and F.

### A. DARKNET

The anonymous networks that enjoy the highest levels of popularity include Tor, I2P, Freenet, Zeronet, and JonDonym. Tor is a widely utilized communication network that prioritizes anonymity, boasting a substantial user base. The protocol is founded upon the Transmission Control Protocol (TCP) and employs a multi-hop technique for establishing communication links. Within a preexisting communication pathway connecting the ingress node and the egress node, the Tor network will employ a stochastic process to designate a set of relay nodes, numbering greater than three, from the directory server. In the relay network, individual nodes only know of their immediate predecessor and successor nodes. The selection of these nodes is determined randomly, and the connections between them undergo continuous fluctuations. Simultaneously, the data within the Tor network is concealed within numerous layers of encryption, and all communications go through a sequential process of encryption at each layer [18].

I2P is an enhanced anonymity network that builds upon the principles of Tor. It eliminates the reliance on a central node and instead employs a fully decentralized architecture, enhancing both the anonymity and stability of the network [60]. A multi-level encrypted tunnel system is employed to obfuscate the identification information and communication relationship of the involved parties.

JonDonym offers users a range of anonymous services by combining hybrid cascades. Each hybrid cascade consists of two or three encrypted hybrid servers. In contrast to Tor, the JonDonym network maintains a static link. The user can choose the link for data transfer but cannot modify the composition of various nodes on the link. The classification of anonymous network traffic has become increasingly difficult due to the implementation of encryption technology [48].

Freenet is a decentralized network that facilitates the public distribution of data. The primary objective is to ensure confidentiality, specifically safeguarding the interests of those who disclose sensitive information or engage in advocacy activities. Therefore, the concealment of the identities of content suppliers and subscribers is implemented to protect them from potential persecution. Anonymity might potentially be exploited by terrorists to launch attacks, acquire influence, and evade law enforcement entities. Freenet is a decentralized network that facilitates anonymous peer-to-peer communication, providing a means for data authors and retrievers to maintain their anonymity. The user designates a portion of their hard disk within the system and utilizes it as a distributed storage system for sharing purposes. The Freenet system determines the privileges of insertion, retrieval, and deletion, while the specific placement of a shared file is decided by a unique routing key linked with it [61]. Therefore, every individual within the Freenet possesses knowledge about their immediate neighboring peers. Furthermore, the incorporation of Freenet's anonymity is achieved by rewriting the source of messages at each peer and hop-by-hop forwarding of user communications.

ZeroNet is a network comprising peer-to-peer users. It operates in a decentralized manner, resembling the structure of the World Wide Web. Instead of utilizing an IP address, websites are distinguished by a public key, specifically a Bitcoin address. These websites can be viewed through a conventional web browser by employing the ZeroNet application. By default, ZeroNet does not provide anonymity; however, it can be configured to utilize the Tor network for routing purposes. Additionally, ZeroNet employs trackers sourced from the BitTorrent network to facilitate the establishment of connections among peers. One of the primary advantages of ZeroNet is its capacity to provide offline site accessibility [62]. Furthermore, the presence of seeders for a particular site renders it impervious to removal from the ZeroNet platform, hence rendering Digital Millennium Copyright Requests for takedown ineffective. ZeroNet has garnered significant popularity among cybercriminals, attracting many extreme and terrorist entities, both in the cyber realm and beyond.

The Java Anon Proxy (JAP), alternatively referred to as JonDonym, was developed as a proxy system to enable individuals to browse the internet while maintaining the ability to revoke their pseudonymous identity. JonDonym does not function as a virtual private network (VPN) service. The anonymizing service functions in a manner akin to the Tor network. The system combines the user's traffic and applies encryption to it. JonDonym is a hybrid cascades network that uses multilayer encryption to give people services that keep them anonymous. On the JonDonym network, the cascades consist of two (free) or three (paid) mix servers. Users can choose the mixed cascades to build the link, but they can't change the cascades. All of these links from different users are combined into one to the first mix server, which is then sent to the next mix in the chain [63].

### B. DARKNET TRAFFIC ANALYSIS PROCESS (RQ2)

The approach employed for analyzing darknet traffic exhibits variability contingent upon the particular aims of the investigation.

The initial step undertaken by researchers is the acquisition of an extensive dataset comprising network traffic occurring within the darknet.

The acquisition of this data can be accomplished through diverse methodologies, including the monitoring of darknet activity or the utilization of network telescopes that are specifically engineered to capture traffic originating from unutilized IP address space. After the collection of the dataset, researchers utilize a range of approaches and tools to conduct its analysis. Various methodologies can be employed, including packet-level analysis, statistical analysis, machine learning algorithms, and deep neural networks. A prevalent strategy involves employing statistical estimating techniques, including maximum likelihood, methods of moments, and linear regression estimators, to deduce temporal dynamics and trends within the darknet traffic. In this section, only the process of the traffic analysis is discussed. However, further sections will be presenting a detailed elaboration on each part of the process.

**Data Collection:** This component describes the data sources, the data collection size, and the dataset's availability for the models. Researchers and businesses use several types of data while collecting information. Many studies have used data already scraped from internet databases; others have used onion sites as their dataset, while some used real-time darknet traffic to achieve the required output. We can safely divide the datasets into public, private and real-time data. Detailed discussion is provided in section C.

The darknet traffic input can be public, private or real-time data, which can be classified into three categories based on its attributes [64].
- **Circuit:** It has all the information on circuit lifetime, cell inter-arrival times, cells per circuit lifetime, uplink cells and the rate of the downlink cells to the uplink cells.
- **Flow:** Flow segment size, round trip time, and duration.
- **Packet:** Gives all the information regarding packet length, frequency, and header.

**Data Pre-Processing:** This is a critical phase in the data processing process. The major components of this process include important feature extraction, data balancing, data filtering, duplication or noise reduction, and feature selection. This stage is usually followed by their appropriate needs to feed the model in most research.

**Data Processing:** This is the essential stage of any model because it involves the implementation of algorithms, such as machine learning or deep learning techniques, data classifications, clustering or labeling, testing the performance of the model with training and testing data, and applying the techniques to their respective fields [65]. More explanation is provided in section D.

**Output:** The outcome aimed to develop the framework is the results, which vary depending on the deployed model. It could be in the form of alerts, reports, graphs, mail notifications, or images.

The traffic analysis process can give different types of outputs based on the purpose of the analysis process. Flow-based analysis provides the traffic classifications of the network browser flows, further packet-level analysis can provide the application type and software used by the particular type, which is a fine-grained output. The protocol-based analysis provides the classification of protocols. However, behavior-based analysis comes up with attack alerts, anomaly detection and sometimes proposals for new attacks or protection against attacks. A detailed discussion of the taxonomy of traffic analysis is provided in Section VI of this paper.
- Traffic Cluster: Tor, I2P, Freenet, JonDonym, VPN, Zeronet.
- Application Type: Video, Audio, Browsing, Email, Chat, VoIP, Torrenting.
- Application Software: Facebook site, YouTube video, Skype call, Windows update.
- Application Protocol: HTTPS, FTP, P2P.
- Behavior Alerts: Malware, Botnet, Attack prediction.

### C. DARKNET DATASETS

Here we presented details for the datasets used for the traffic analysis input as this is the most crucial element in the traffic analysis, which leads to the feature and classifier selection and classification of traffic.

TABLE VI gives the details of the dataset attributes and Figure 10 gives the percentage of its usage in the selected literature.

| TABLE VI Summary of Publicly Available Darknet Datasets | | | | | |
|---|---|---|---|---|---|
| **Dataset** | **Tool** | **Type of Traffic** | **Instances** | **Used By** | **Available Link** |
| ANON17 | Tor | Normal Tor Traffic | 5,283 | [63][18][22] | https://projects.cs.dal.ca/projectx/Download.html |
| | | Application On Tor | 252 | | |
| | | Tor Pluggable Transport | 353,391 | | |
| | I2P | I2P Applications Tunnels with other Tunnels – 80% Bandwidth | 449,987 | | |
| | | I2P Applications Tunnels with other Tunnels – I2PApp0BW 0% Bandwidth | 195,081 | | |
| | | I2P User Traffic | 449,998 | | |
| | | I2P Application | 640 | | |
| | JonDonym | JonDonym Traffic | 5,440 | | |
| Darknet-Dataset-2020 | Tor | Tor Application | 8,632 | [66][67] | https://github.com/huyz97/darknet-dataset-2020 |
| | I2P | I2P Application | 8,148 | | |
| | Freenet | Freenet Application | 16,387 | | |
| | Zeronet | Zeronet Application | 15,477 | | |
| CIC-Darknet2020 | Non-Tor | Non-Tor Application | 93,357 | [12][68][57][30][9][31][10][8][38][69][15][70][13][56][28][71][26] | https://www.unb.ca/cic/datasets/darknet2020.html |
| | Non-VPN | Non-VPN | 23,864 | | |
| | Tor | Tor Application | 13,933 | | |
| | VPN | VPN application | 22,920 | | |
| SJTU-AN21 | I2P | I2P Application | 26,957 | [20] | https://github.com/iZRJ/The-SJTU-AN21-Dataset |
| | Tor | Tor Application | 4.282 | | |
| | JonDonym | JonDonym Traffic | 2.221 | | |
| UNB-CIC | Tor | Tor Traffic | 1,415,371 | [72][14][32] | https://www.unb.ca/cic/datasets/tor.html |
| | | Tor Application | 2,830,743 | | |
| | | Non-Tor Traffic | 1,415,372 | | |

### 1) PUBLIC DATASETS

**Anon17:** Anon17 was obtained from the NIMS lab during 2014–2017. The data collection took place in an authentic network setting, as shown by reference [63].

The dataset comprised three distinct anonymity networks – Tor, I2P and JonDonym. Multiple obfuscation techniques are employed within the Tor network, and various apps are utilized across these anonymity networks. The researchers utilized Tranalyzer to extract a comprehensive set of 1,010,962 flows from the PCAP files. These flows encompassed 82 distinct properties, including time (representing the time of each flow), maxPL (indicating the maximum packet length within a flow), and Dir (denoting the direction of the flow) [18][22].

**CIC-Darknet2020:** The dataset known as CIC-Darknet2020, as described, is a compilation of two publicly available datasets from the University of New Brunswick [12]. The study integrates the ISCXTor2016 and ISCXVPN2016 datasets, which were collected using Wireshark and TCPdump to capture real-time network traffic [12]. The CICFlowMeter, developed by Lashkari in 2018, is employed to extract the properties of the CIC-Darknet2020 dataset from the provided traffic samples. The CIC-Darknet2020 dataset comprises samples that contain traffic features obtained by extracting relevant information from raw traffic packet capture sessions. The dataset known as CIC-Darknet2020 comprises a total of 141530 samples and 85 features. The highest-level traffic category labels encompass Tor, non-Tor, VPN, and non-VPN. Within these high-level categories, samples are further classified based on the sorts of applications employed to create the traffic. The aforementioned subcategories encompass audio-streaming, browsing, chat, email, file transfer, peer-to-peer (P2P) communication, video streaming, and voice-over-internet protocol (VoIP) services.

**PTO dataset:** The PTO dataset, publicly introduced by Petagna et al. in 2019, consists of Tor application traffic data. It lasts around four hours and encompasses the Tor traffic generated by ten Android applications. The writers apply automated scripts to manipulate a smartphone's functionality to generate traffic, and they utilize Orbot to proxy the traffic to the Tor network. The router will capture and store the traffic traveling through it into a local dataset. Furthermore, the authors have included two distinct datasets in their study – connection padding and reduced connection padding methodologies. The concept of connection padding refers to the automatic transmission of a padding cell to counteract traffic analysis when a predetermined time limit is exceeded by a timer. The presence of padding can result in increased overhead, thereby decreasing the frequency of exchanging padding cells when connection padding is reduced. The connection padding dataset was selected because of the prevalence of connection padding as the default setting in Orbot [17].

**UNB-CIC:** The UNB-CIC Tor Network Traffic dataset is an exemplary dataset that captures real-world network traffic and consists of a collection of tasks. A total of three users were designated for collecting browser traffic data, while two users were assigned to handle communication activities, such as chat, mail, and peer-to-peer interactions. These activities were conducted across a selection of over 18 widely-used programs, including Facebook, Skype, Spotify, and Gmail, among others. The dataset encompasses eight distinct categories of Tor traffic alongside non-Tor traffic. It includes non-Tor traffic with distinct characteristics that distinguish it from Tor traffic. These attributes are sometimes referred to as features. The dataset pertaining to the UNB-CIC Tor Network Traffic comprises a comprehensive set of 28 distinct features [72].

The features were derived from packets that shared identical values for the source IP, source port, destination port, and protocol (TCP and UDP). The entire Tor network traffic is transmitted via the Transmission Control Protocol (TCP) due to the absence of support for the User Datagram Protocol (UDP). The formation of flows was facilitated by utilizing a novel application known as the ISCX Flow Meter, which effectively produces bidirectional flow ID [14][32].

**Darknet-Dataset-2020:** This dataset was developed in 2020 by Youzong et al. [6] and contains 48,644 instances and 26 characteristics. Eight different types of user behavior traffic were collected (Browsing, Chat, E-mail, Audio-streaming, Video-streaming, File Transfer, P2P, and VoIP) in Tor, I2P, ZeroNet, and Freenet and the resulting dataset was made publicly available online.

**SJTU-AN21:** Version 0.4.1.5 of the Tor network, implemented in 2018, introduced notable modifications to enhance user anonymity by mitigating the risk of deanonymization through traffic analysis. These alterations encompass the inclusion of circuit-level padding and SENDME units. In 2019, version 0.9.36 of the I2P network incorporated the implementation of the NTCP2 protocol. The NTCP2 protocol leverages the Noise protocol framework to enhance its resistance against Deep Packet Inspection (DPI) attacks. Nevertheless, the two preceding datasets pertaining to anonymity networks were made available before 2017. To adapt the classifier for use in contemporary anonymity network traffic analysis, R. Zhao et al. gathered traffic data from ten anonymity services in the most recent iterations of the three prominent anonymity networks, namely Tor, I2P, and JonDonym. The SJTU-AN21 dataset encompasses four primary anonymity services inside the I2P network, namely Eepsites, IRC, Snark, and Video. Conversely, the Tor network comprises five principal anonymity services, namely BitTorrent, Chat, FTP, Streaming, and Browsing [20].

**Alexa Dataset:** Most studies on website fingerprinting have relied on the utilization of the Alexa Top Ranked list, which is a compilation of the most frequently visited URLs on the internet as determined by the web analytics service Alexa, which we also use to assess the efficacy of our fingerprinting methodology in real-world settings [73].

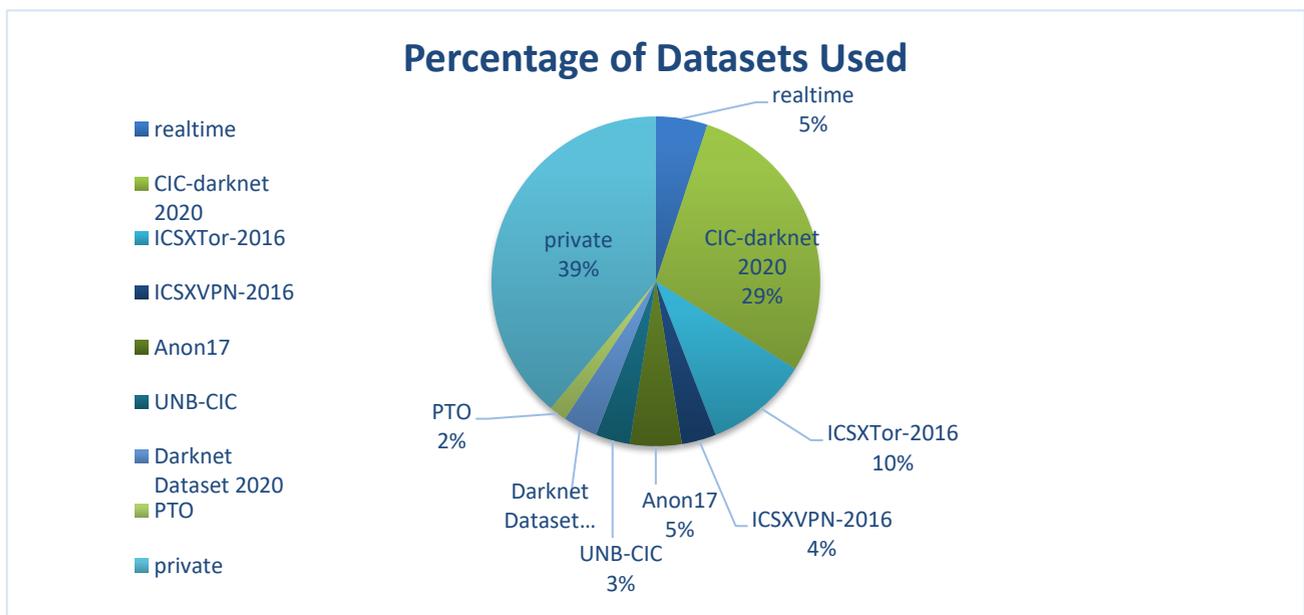

*Figure 10 Classification of Dataset Used*

2) PRIVATE DATASET

Dodia obtained data through a malware corpus sourced from VirusTotal (VT), a widely used platform for exchanging threat information among numerous security researchers and many security organizations. The VT platform compiles detection outcomes from a comprehensive range of 72 distinct AntiVirus (AV) engines [52].

Whereas Ban et al. [45] used the data privetly gathered on their project called NICTER, which is developed on data-mining engines. Their microscopic component employs honey pots and email-traps to capture malware programs in their natural environment. The programs gathered are input into a behavior analyzer and a code analyzer to acquire profiles of their characteristics and behaviors. This process is conducted to detect Botnets and DDos attacks.

Gioacchini established a darknet within the IP range of a university campus network, specifically utilizing a /24 subnet to conduct experimental activities. The dataset consists of 30 days of data, with the initial 29 days being utilized for training algorithms, while the final day is reserved as an independent test set. The darknet can monitor many unique sender IP addresses, which collectively transmit tens of millions of packets. The researchers monitor packets transmitted across all ports, noting a significantly imbalanced distribution wherein a considerable proportion of packets are directed toward the most often targeted ports. This pattern is observed across many sources, numbering in the tens of thousands [39][74].

3) REAL-TIME DATA

The GLASSO engine, developed by Han et al. [75] in 2016 operates in batch mode and necessitates a three-day collection of darknet data for processing. Consequently, the analysis outcome cannot be obtained in real-time and may be subject to a delay beyond three days in the most unfavorable scenario. The researchers have developed a methodology that facilitates the real-time detection of emerging malware activity, hence enhancing the ability to respond to such incidents promptly. Several evaluations have been undertaken to illustrate the efficacy of the engine. The authors clarified that the engine generated alerts for three distinct categories of activity, namely cyberattacks, survey scans, and sporadically focused traffic. They emphasized the importance of establishing appropriate parameters to effectively differentiate between these activities.

D. FEATURE SELECTION (RQ2.2)

In the field of machine learning, the acquisition of substantial quantities of data is a common practice aimed at enhancing the algorithm's training process. However, it is frequently impractical to handle such extensive datasets. Feature selection refers to identifying and choosing pertinent features or a subset of features. Evaluation criteria are employed to identify an optimal subset of features. Feature selection algorithms exhibit distinct characteristics, namely search organization, generation of successors, and evaluation measures. The assessment process employed by FS incorporates various factors, including probability of error, divergence, dependence, interclass distance, information gain, and consistency.

The process of feature selection is of utmost importance in the identification and analysis of darknet traffic, as it aids in the identification and characterization of the most effective feature set. The approach consists of two main steps: firstly, the pre-processing of the darknet-dataset to extract features and determine target labels; and secondly, the selection of the important features. During the data pre-processing phase, data balancing noise cancellation and feature extraction through CICFlowMeter or Tranzylyzers are utilized. After selecting the most impacting features as per the respective model, data is forwarded for further processing either for data balancing or directly by machine learning or deep learning techniques. The data balancing step can be performed before or after feature selection.

Data balancing plays an important role in overall output accuracy. A popular data balancing method is the Synthetic Minority Oversampling Technique (SMOTE). This process aims to address the issue of imbalanced class set sizes within the traffic datasets, hence mitigating any classification biases. The SMOTE technique, as described in reference [56], generates a balanced dataset by artificially augmenting the number of samples belonging to the minority class in an imbalanced dataset. This methodology is commonly employed in areas characterized by limited accessibility or high costs associated with data acquisition, particularly in domains such as healthcare and internet traffic analysis. This approach effectively mitigates the issue of excessive data allocation and yields satisfactory results in terms of classification accuracy. In the SMOTE algorithm, the temporal aspect of a synthetic instance is introduced.

Sridhar et al. [57] introduced the technique of oversampling minority class instances implemented using generative adversarial networks (GANs). The conditional GAN is employed for generating examples that belong to a specific class. Instead of generating arbitrary samples of traffic data, data related to a specific class is generated by including the class label as a feature in both the discriminator and generator components. He then applied the Chi-Squared algorithm to effectively classify instances using feature importance and identify the impact of each feature on the classification process. It helps to determine which features are independent and have little impact on the classification process.

The process of selecting features plays a pivotal role in the analysis of darknet data. The accuracy of encrypted traffic categorization can be greatly enhanced by using optimal feature selection, hence aiding in identifying unlawful activity within the darknet. Nevertheless, identifying the most important aspects is challenging due to the vast volume and intricate nature of darknet data. Hence, sophisticated machine learning methods are frequently utilized to address this concern. A comprehensive understanding of these characteristics can additionally aid in enhancing existing methodologies and developing novel strategies for the efficacious investigation of darknet traffic. In addition, the feature selection process is crucial in minimizing the computational and time complexity involved in the analysis, enabling models to concentrate on the traits most relevant to darknet activity. Conversely, inadequate feature selection can result in overfitting and erroneous findings. Therefore, it is crucial to incorporate a precise and fast feature selection methodology as an essential component of dependable and impactful analysis of darknet traffic.

During the pre-processing stage of a machine learning (ML) pipeline, feature selection methods are employed to exclude attributes that are deemed irrelevant. In general, feature selection algorithms can be classified into three main groups: (a) filter, (b) wrapper, and (c) hybrid techniques. Filter approaches utilize inherent features of attributes, such as distance, entropy, dependency, or consistency, to ascertain whether a subset of attributes adequately represents the characteristics of observations. Wrapper techniques employ the accuracies of machine learning algorithms to assess the significance of chosen features. Nevertheless, the results obtained from the chosen subset of variables tend to exhibit a bias toward the machine learning technique that is utilized. Hybrid methodologies integrate the advantages of both wrapper and filter techniques to identify an optimal subset of attributes. In this research, the Correlation Based Filter Selection (CFS) and Symmetric Uncertainty (SU) feature selection methods were employed to acquire the desired outcomes [14].

Liukun et al. [17] introduce FlowMFD, an innovative method for classifying application traffic based on Tor. FlowMFD leverages chromatographic features, specifically amount-frequency-direction (MFD), along with spatial-temporal modeling techniques. The FlowMFD system examines the interaction behavior between Tor apps and servers by analyzing time series features (TSFs) associated with packets of varying sizes. The MFD chromatographic features (MFDCF) are specifically intended to depict the pattern accurately. The aforementioned characteristics facilitate the amalgamation of numerous low-dimensional time series features (TSFs) onto a singular plane while effectively preserving most pattern information.

| | TABLE VII Feature Selection Algorithms in the Darknet Traffic Analysis | | | |
|---|---|---|---|---|
| **Study** | **Dataset** | **Feature Selection Algorithm** | **Technique** | **Evaluation Attributes** |
| [27] | CIC-darknet | Amount-frequency-direction (MFD) | RF, KNN, SVM | FN, FP, F1, AUC, acc, precision, recall |
| [76] | CIC-darknet | Analysis of Variance (ANOVA) | RF | TP, TN, FP, FN, Precision, recall, f1, acc, kappa, MCC, confusion matrix |
| [14] | UNBCIC | Correlation Based Filter Selection (CFS) and Symmetric Uncertainty (SU) | DNN | acc, F1, precision recall, ROC, AUC, |
| [18] | Anon17 | MMIRF (MMI+RF) | XGB | Precision, recall, f-measure |
| [58] | CIC-darknet | N-gram Recursive Feature Elimination technique | DT, RF | Precision, recall, f1, confusion matrix |
| [77] | Private | Compressed traffic features | SVM, KNN | acc |

| [41] | Private | FastText for feature extraction | DBSCAN | N/a |

Coutinho and colleagues conducted feature extraction and employed an n-gram technique to categorize potential subnets. In addition, the researchers assessed the significance of the optimal features chosen by the recursive feature elimination technique in relation to the given problem. The utilization of n-gram models presents a more comprehensive mapping strategy for the encoding of IP addresses. Originally, these models were suggested for natural language processing and presently they hold a prominent position as the prevailing representation in several detection systems. The utilization of n-grams can reduce the occurrence of false positives in predictive models. Therefore, the dataset was expanded by utilizing the IP addresses and breaking them into individual unigrams, bi-grams, and trigrams.

The concept of compressive traffic analysis was introduced by Nasr et al. [59], who suggested using compressed traffic features for conducting traffic analysis activities, instead of using raw traffic features. Consequently, using fewer features leads to enhanced efficiency in terms of storage, communications, and compute overheads associated with traffic analysis. Compressive traffic analysis is a prominent field in signal processing that utilizes linear projection algorithms to compress traffic data effectively. Lately, the utilization of compressed sensing has been employed in the context of networking issues, specifically pertaining to the estimate and completeness of network datasets and network traffic matrices.

Ishikawa et al. [41] employed FastText for feature extraction to investigate the inherent correlation between targeted network services, as inferred from the destination ports of scanning packets. Subsequently, a nonlinear dimension reduction technique known as UMAP is utilized to map hosts onto a two-dimensional embedding space, to facilitate visualization. Ultimately, clustering analysis is conducted utilizing the DBSCAN algorithm to autonomously detect clusters of compromised hosts exhibiting comparable attack patterns.

The primary objective of feature selection is to minimize the number of features employed for classification by eliminating redundant characteristics within the dataset. The determination of feature exclusion criteria can be accomplished by either unsupervised methods, such as variance and correlation thresholds, or supervised methods, such as univariate statistical tests. TABLE VII gives an overview of feature selection algorithms used in the traffic analysis.

### D. DATA PROCESSING APPROACHES (RQ2.3)

Machine learning in traffic monitoring encompasses three primary approaches: supervised, semi-supervised, and unsupervised. In traffic categorization, the utilization of various approaches yields varying outcomes, with the effectiveness and dependability contingent upon the specific objectives and dataset employed as input.

Figure 11. Presenting all the techniques used in the analysis process of darknet traffic.

#### 1) SUPERVISED LEARNING

It used pre-training datasets, which consist of labeled data that are categorized according to specific traffic criteria, to train the algorithms for traffic classification. Some examples of supervised learning algorithms include K-Nearest Neighbors (k-NN), Bayesian Network, Decision Tree, and Support Vector Machine (SVM). One primary benefit of this approach is its ability to achieve a low percentage of incorrect categorization. However, there may be difficulties in obtaining comprehensive sets of pre-training data and it may necessitate a lengthy training period.

**Classification**: In classification tasks, the machine learning program must draw a conclusion from observed values and determine to what category new observations belong. There are four different types of Classification Tasks in Machine Learning and they are the following.

      Binary Classification
      Multi-Class Classification
      Multi-Label Classification
      Imbalanced Classification.

**Regression**: In regression tasks, the machine learning program must estimate – and understand – the relationships among variables. Regression analysis focuses on one dependent variable and a series of other changing variables – making it particularly useful for prediction and forecasting.

**Forecasting**: Forecasting is the process of making predictions based on past and present data and is commonly used to analyze trends.

2) SEMI-SUPERVISED LEARNING

It shares similarities with supervised learning, but it differs in that it does not rely exclusively on fully labeled training data. The pre-training data comprised solely of partially labeled data. One primary benefit is the reduced requirement for many pre-training data sets. However, the issue of accuracy may be a significant challenge, particularly when classifying different types of encrypted application communication.

*Reinforced Learning*: Reinforcement learning focuses on regimented learning processes, where a machine learning algorithm is provided with a set of actions, parameters, and end values. By defining the rules, the machine learning algorithm then explores different options and possibilities, monitoring and evaluating each result to determine which one is optimal. Reinforcement learning teaches the machine trial and error. It learns from past experiences and begins to adapt its approach in response to the situation to achieve the best possible result.

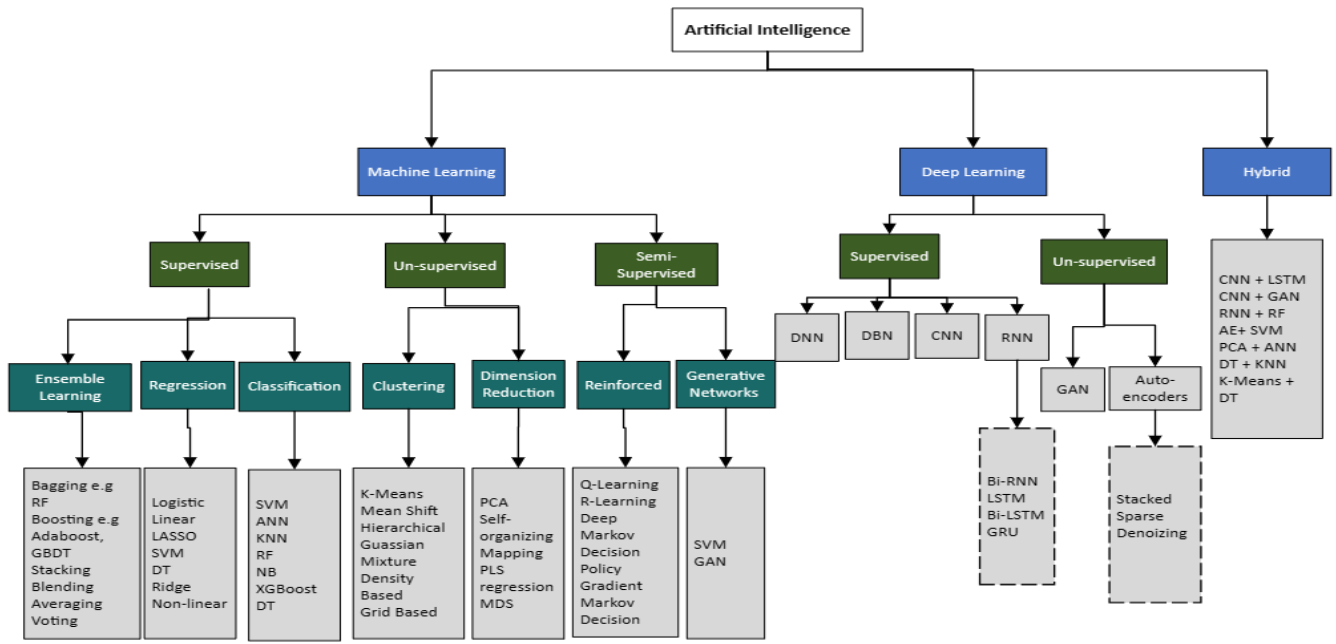

*Figure 11 Learning Techniques used for Traffic Analysis*

3) UNSUPERVISED LEARNING

This technique is primarily utilized for data clustering, without the reliance on any pre-training data. Two examples of unsupervised learning algorithms are Fuzzy C-means and K-means. Although it does not require training data and is easy to use, it is more typically employed for the process of traffic clustering rather than for the categorization of encrypted traffic. One of the primary applications of unsupervised traffic clustering is anomaly detection, which demonstrates efficacy in analyzing unlabeled traffic data. Clustering and dimension reduction fall under the umbrella of unsupervised learning.

**Clustering**: Clustering involves grouping sets of similar data (based on defined criteria). It's useful for segmenting data into several groups and analyzing each dataset to find patterns.

**Dimension reduction**: Dimension reduction reduces the number of variables being considered to find the required information.

4) DEEP LEARNING

Deep learning utilizes artificial neural networks to do complex computations on vast quantities of data. A neural network is organized in a manner that resembles the anatomical structure of the human brain, comprising artificial neurons, sometimes called nodes. The nodes are arranged in a stacked configuration, with each layer consisting of adjacent nodes. The input layer refers to the initial layer of a neural network model where the input data is received and processed. The hidden layer(s) and the final layer in the neural network architecture, commonly called the output layer, produce the desired predictions or classifications based on the input data and the learning model There are a number of deep learning algorithms like Convolutional Neural Networks, resNet50, etc. We plan to explore various deep learning algorithms to train and test.

### E. EVALUATION MATRIX (RQ2.4)

When assessing a machine learning model, it is crucial to select metrics that are in accordance with the particular objectives and demands of the work at hand. In a traffic analysis task, the prioritization of recall above precision may be more significant. Furthermore, it is essential to consider using metrics in various scenarios, such as multiclass classification, regression, and other specialized tasks. This analysis presents the distribution of evaluation attributes in terms of their respective usage percentages. Following are some attributes of the evaluation matrix that are mostly used in the reviewed papers. Figure 12 gives the percentage usage of these attributes in the selected literature.

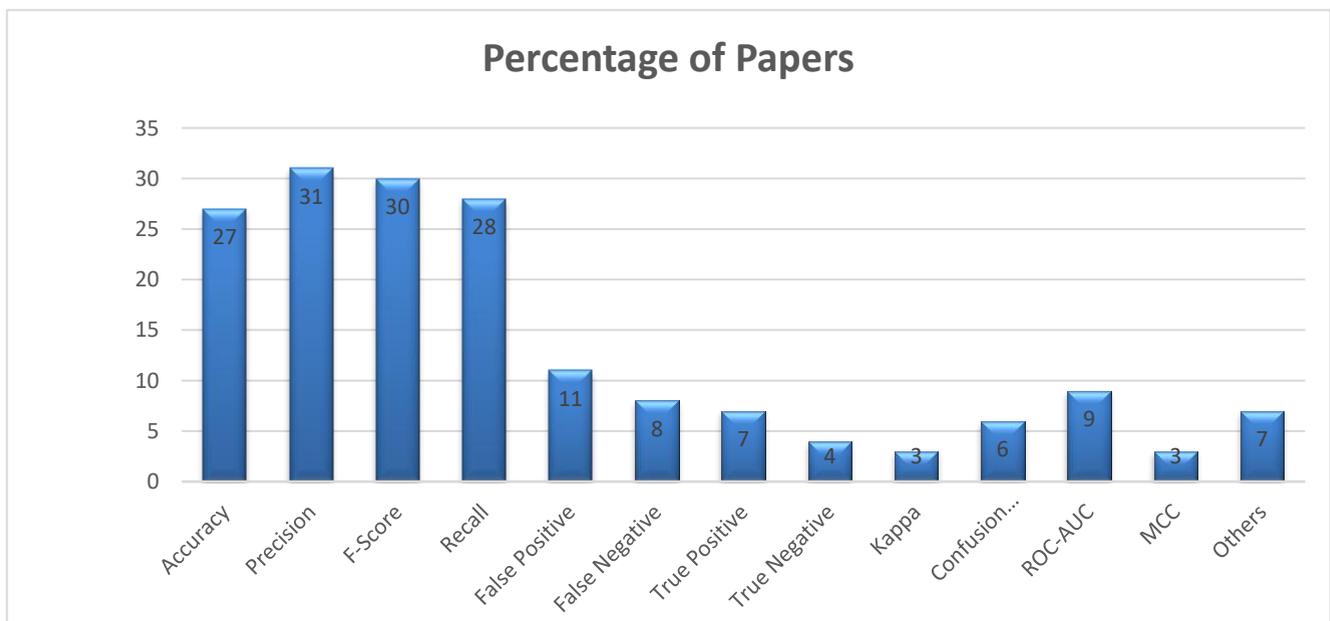

*Figure 12 Usage of Evaluation Attributes*

- Accuracy refers to the ratio of accurately classified cases to the total number of instances. The statistic in question is fundamental in nature, although its applicability may be limited when dealing with skewed datasets.
- The precision metric quantifies the proportion of correctly predicted positive instances among all instances that were anticipated as positive.
- The recall metric quantifies the proportion of accurately anticipated instances belonging to the positive class. The true positive rate, also known as sensitivity or recall, is defined as the ratio of correctly predicted positive instances to the total number of actual positive instances. The metric quantifies the model's capacity to accurately detect all pertinent events.
- The F1 Score can be defined as the mathematical average of precision and recall, where the harmonic mean is used as the combining function. The aforementioned approach achieves a harmonious equilibrium between precision and recall.
- The Area Under the Receiver Operating Characteristic (ROC-AUC) is a quantitative measure that takes into account the balance between the genuine positive rate and false positive rate at various threshold levels. The Receiver Operating Characteristic (ROC) graph is a widely used method in the field of machine learning for visually representing, categorizing, and choosing classifiers according to their effectiveness. ROC graphs are widely used in the machine learning field because they provide a more robust method for evaluating performance compared to basic classification accuracy, which is sometimes an inadequate metric. When examining classification issues that include only two classes, it is important to note that there are four potential outcomes when evaluating a classifier and an instance.

- A true positive (TP) refers to an event that is classed as positive when it is indeed positive.
- A false negative (FN) occurs when an incident that is actually positive is incorrectly labeled as negative.
- A true negative (TN): The instance exhibits a negative condition and is correctly classed as negative.
- A false positive (FP) occurs when an incident that is actually negative is incorrectly labeled as positive.
- The confusion matrix is a performance evaluation tool that facilitates the graphical representation of the effectiveness of an algorithm, typically employed in the context of supervised learning algorithms. In the context of multiclass classification, the confusion matrix is a matrix of dimensions N × N, where N represents the total number of classes. The confusion matrix provides a clearer representation of true positives (TP), true negatives (TN), false positives (FP), and false negatives (FN).
- The Matthews Correlation Coefficient (MCC) is a statistical measure that quantifies the correlation between the observed and anticipated binary classifications. The measurement scale ranges from -1 to 1, where a value of 1 signifies accurate and precise predictions.

## VI. TAXONOMY OF THE DARKNET TRAFFIC CLASSIFICATION (RQ3)

This section discusses the architectural analysis of the literature reviewed to detect threats. We categorized the literature into four broad categories based on the detection target: 1. Flow-based detection (detection of anonymity tools) 2. Packet-based detection 3. Protocol-based detection and 4. Behavior-based detection. We employed a generally used process to summarize the detecting architecture.

Darknet traffic can be classified as the detection of the darknet protocol. It also depends on the database the researcher used to achieve the classification target. In TABLES VIII, IX, and X, we present studies and the targeted traffic whereas in TABLE IX we discuss darknet traffic classification techniques as per their targeted traffic and output.

### A) FLOW-BASED TECHNIQUES

The identification and analysis of darknet traffic provide an important obstacle for researchers. Flow-based techniques have been investigated to analyze network traffic in the darknet to detect traffic of Tor or other anonymous tools. The mentioned methodologies prioritize the analysis of flow-level characteristics, including packet size, inter-arrival time, and flow duration, to differentiate between Tor and non-Tor network traffic. The importance of precise and efficient detection techniques for the comprehensive analysis of darknet traffic is underscored by an extensive assessment of the literature. Identifying and defending against cyber crimes in the darknet is of utmost importance. Flow-based techniques are used to detect tools used in the darknet. It can also be used along with the payload level classification to further categorize the application type used in the particular browser. Figure 13 gives a detailed overview of the darknet traffic classification techniques. We will further classify flow-based techniques as the detection of Tor, I2P, Freenet, Zeronet, and Jondonym. TABLE VIII presents flow-based techniques.

#### I. DETECTION OF NON-OBFUSCATED TOR TRAFFIC

Detection of Tor traffic can be further categorized as detection of obfuscated Tor traffic and non-obfuscated Tor traffic. Khalid Shahbar et al. [22] devised a classifier that utilizes a machine-learning model to analyze both circuit-level and flow-level data. The classification of circuits at the circuit-level considers factors such as the lifespan of the circuit, and the cell rate for both uplink and downlink communication and achieves a level of accuracy exceeding 94%. Tranalyzer2 and Tcptrace are utilized to obtain flow-level samples, encompassing packet length, Inter Arrival Time, and packet inter distance as their distinguishing characteristics. Additionally, the authors presented a proposed model in their study that aims to detect Tor pluggable transports. This model achieved an accuracy of 94% by analyzing the background traffic and considering five specific features: Source-Destination IP, Source Port, Destination Port, and Protocol.

Lingyu et al. [78] introduced a hierarchical classification approach that utilizes the decision tree algorithm for Tor traffic identification and the Tri-Training algorithm for Tor traffic segmentation. The Tri-Training method is a machine learning algorithm categorized as semi-supervised learning. It makes use of the co-training technique. One of the advantages of this approach is that it necessitates a smaller amount of training data compared to supervised approaches. Additionally, it does not necessitate cross-validation or impose restrictions on the base classifier. This study primarily examined packet-based properties, including packet length entropy, frequency of 600-byte packets, frequency of zero data packets (first 10), and average packet interval duration. The outcome demonstrates a notable level of accuracy in categorization, which may be attributed to the utilization of a hierarchical instrument.

Alimoradi et al. [9] suggested a novel decision support system called a Tor-VPN detector to categorize raw darknet data. The detector uses a deep neural network design with 79 input artificial neurons and six hidden layers to uncover complicated nonlinear relations from raw darknet traffic. Analyses are performed on a DIDarknet benchmark dataset to assess the efficacy of the proposed technique. With a 96% accuracy rate, the model is superior to the current gold-standard neural network for classifying darknet traffic. Our model's effectiveness in dealing with darknet data is demonstrated by these findings without the need for any pre-processing approaches, such as feature extraction or balancing methods.

In their study, AlSabah tried to categorize Tor traffic based on the specific application being utilized. The study's authors included interactive online surfing and bulk downloading, specifically focusing on BitTorrent and streaming applications as the sorts of applications under consideration in their research. The researchers determined that the act of bulk downloading consumes a significant amount of bandwidth, albeit accounting for a negligible proportion of the overall number of connections. The authors' objective was to offer distinct levels of Quality of Service (QoS) to various categories of traffic. The researchers employed many metrics like cell Inter-Arrival Times (IAT), circuit longevity, and volume of data transmitted upstream and downstream, as well as classification techniques such as Naïve Bayes, Bayesian Networks, and Decision Trees. The experimental findings presented in reference [2] demonstrate a classification accuracy exceeding 95% in identifying application types for Tor circuits within an operational Tor network [79].

Karunanayake et al. [6] conducted a study to determine how Onion service traffic could be classified. Their research focused on three primary contributions. Initially, the researchers endeavor to discern Onion Service traffic from other forms of Tor network traffic. The employed methodologies can accurately discern Onion Service traffic with a precision above 99%. However, the researchers assessed the performance of their convolutional neural network (CNN) techniques when subjected to alterations in Tor traffic. The empirical findings indicate that under these circumstances, the discernibility of Onion Service traffic is reduced, with a decrease in accuracy exceeding 15% observed in certain instances.

Deep Learning (DL) is a subfield within the broader domain of machine learning, and it finds extensive use in various domains, including image classification and speech recognition. In general, algorithms that incorporate artificial neural networks (ANNs) are classified in this category. In their study, Kim et al. [32] employed Convolutional Neural Networks (CNN) to categorize Tor traffic compared to non-Tor data. The researchers employed hexadecimal raw packets in conjunction with a convolutional neural network (CNN) to achieve a comprehensive accuracy of 99.3% in classifying several application categories.

## II. DETECTION OF OBFUSCATED TOR TRAFFIC

To enhance its dependability and anonymity, Tor employs obfuscation techniques to respond to the increasing number of ways that have been devised to identify non-obfuscated Tor traffic. Each of these obfuscation strategies employs different mechanisms to safeguard Tor traffic against detection. The three officially supported plugins are Meek-based, FTE-based, and Obfs4-based obfuscation, as previously stated. In 2018, Yao conducted research on the subject of obfuscated Tor traffic that was based on the concept of meek. Subsequently, in 2019, Cai et al. employed flow analysis techniques to detect and classify the pluggable transport technologies utilized within the Tor network. The researchers presented the outcomes of their classification of pluggable transport technologies utilizing the isAnon model. The findings indicate that it is feasible to classify various obfuscation strategies employed by Tor, achieving an overall accuracy rate of up to 99.91%. In their study, Xu et al. [2] employ a methodology to identify Tor traffic by gathering three distinct types of obfuscated Tor traffic. They subsequently utilize a sliding window approach to extract 12 features from the data stream based on the five-tuple, which encompasses packet length, packet arrival time interval, and the ratio of bytes sent and received. Lastly, the researchers employed XGBoost, Random Forest, and other machine learning techniques to discern obscured Tor network traffic and categorize its different forms. The research conducted by the authors presents a practical approach to mitigate the challenges posed by obfuscated Tor networks. Their methodology successfully identifies three distinct types of obfuscated Tor communication, yielding an impressive precision and recall rate of approximately 99%.

**Meek-Based Obfuscation:** The efficacy of Meek-based obfuscation is attributed to the implementation of domain front technology, wherein diverse domain names are employed across several communication tiers and tunnels to circumvent censorship measures. The communication process involves three distinct entities: the Tor client equipped with the Meek plugin, the fronted server with a domain name authorized by the cloud service provider, and the Tor server also equipped with the Meek plugin. When a user intends to establish a connection with the Tor network, the user's request is enclosed within a Transport Layer Security (TLS) layer. This TLS layer includes the domain name of the fronted server in its header. Subsequently, the user transmits this encapsulated request to the fronted server. Once the fronted server has received the packet, it unpacks the internal request and transmits it to the Tor server. Given the restriction on the server's ability to actively transmit data to the client, the client must engage in continual polling of the frontend server. This polling verifies if the Tor server has transmitted any data in response, ultimately leading to the acquisition of the corresponding response content. In summary, Meek-based obfuscation employs a cloud server with a permissible domain name to redirect queries to the Tor network, thus evading censorship. Consequently, the sent data appears indistinguishable from regular cloud service traffic. Utilizing a polling method leads to the generation of a significant quantity of shorter packets during the communication process, hence manifesting as a conspicuous attribute.

In 2018, Yao et al. introduced a new traffic identification model called MGHMM. This study introduces a novel approach that combines the Mixture of Gaussians (MOG) model with the Hidden Markov Model (HMM) framework for traffic detection. The proposed model utilizes a Mixture of Gaussians (MOG) approach to represent the probability density of two-dimensional observations for each state. Subsequently, the Hidden Markov Model (HMM) is constructed by incorporating the state transitions of traffic during communication. The MGHMM model eliminates PT version restrictions, rendering it a more universally applicable traffic identification model. The researchers utilized actual Tor traffic data obtained from the internet to validate and assess the efficacy and precision of the proposed MGHMM model. The experimental findings indicate that the MGHMM model achieves a high identification rate of 99.4%.

**Obfs4-Based Obfuscation:** The most recent addition to the Obfs proxy, known as Obfs4, employs encryption methods to conceal Tor communication by presenting it as regular encrypted traffic, such as the SSL/TLS protocol. The primary methodology involves the utilization of elliptic curve cryptography (ECC) for data encryption, accompanied by the randomization of payload content. This randomization alters the packet size and effectively masks any characteristics related to packet length inside the network flows. Following the application of random packet length padding, the intended recipient possessing the appropriate key can deduce the accurate packet length value and then reconstruct the packet with precision.

**FTE-Based Obfuscation:** The fundamental concept of FTE-based obfuscation involves the utilization of regular expressions to substitute the bytes present in Tor traffic. One potential use is utilizing a regular phrase that encompasses HTTP protocol keywords. This approach enables the concealment of Tor traffic within HTTP traffic, deceiving the DPI system. Nevertheless, there has been no substantial alteration in the attributes of flows.

III. DETECTION OF I2P TRAFFIC

The isAnon model developed by Cai et al. significantly accelerates learning through the use of parallel and distributed computing. The model's importance lies in the speed with which it can eliminate unnecessary details. By fusing the Modified Mutual Information algorithm with the Random Forest method, the isAnon model creates a cutting-edge hybrid feature selection approach. To avoid overfitting, the proposed model employs a tiered cross-validation strategy that combines an inner 5-fold cross-validation with an outer Monte Carlo cross-validation. The isAnon model has a 99.73% success rate in identifying anonymity networks like Tor, I2P, and JonDonym.

IV. DETECTION OF FREENET TRAFFIC

To differentiate between regular internet traffic and FreeNet traffic, as well as five FreeNet user behaviors, Shi et al. [67] offer a hierarchical categorization approach for FreeNet's network traffic. The weighted K-NN is used to train the classifier on the datset by [62]. The testing results reveal that the suggested classifier can differentiate between regular traffic and FreeNet traffic with an average accuracy of 99.6%, and that it can also distinguish between five user behaviors with an average accuracy of 95.6%. They examined their classifier next to decision trees, Gaussian Naïve Bayes, and K-Nearest Neighbors. According to the findings, the classifier performs best when differentiating between user actions. The classifier's accuracy increases by 1.86%, 57.95%, and 3.10% when compared to the aforementioned three models.

V. DETECTION OF ZERONET TRAFFIC

Yuzong et al. [62] investigated how traffic in Tor, I2P, ZeroNet, and Freenet can be categorized based on user behavior. Due to the enormous number of expected categories in this darknet scenario, they suggest a hierarchical classification approach to identifying darknet user behavior. The trial results demonstrate that the approach has a 96.9% success rate in identifying four distinct darknet kinds and a 92.46% success rate in identifying 25 distinct user behaviors. Six different machine learning methods (LR, DT, RF, GBDT, XGBoost, LightGBM), as well as two deep learning algorithms (MLP, LSTM), were investigated during the training process of the hierarchical classifier to determine which is best suited for the darknet traffic scenario. The results demonstrate that when feature extraction accurately represents traffic characteristics, ML classifiers outperform DL classifiers.

| | | | | TABLE VIII Summary of Flow-Based Traffic Analysis | | | |
|---|---|---|---|---|---|---|---|
| **Authors** | **Year** | **Traffic Attributes** | **Machine Learning Approaches** | **Technique** | **Data Type** | **Features** | **Output Features** |
| Montieri et al. [19] | 2020 | Flow, Packet | Supervised | Naïve Bayes, Bayesian networks, C4.5, RF | Anon17 | 74 features | Classified Tor, I2P and JonDonym with the 75.66% f-measure |
| Cuzzcorea et al. [34] | 2017 | Flow, Packet | Supervised | J48, jRIP, REPTree | Real-world data using tcpdump | 23 features | Detected Tor traffic with the precision equal to 1 |
| Lashkari et al. [12] | 2020 | Flow, Packet | Supervised | CNN | Merged two public datasets ISCXTor2016, ISCXVPN2016 and comes up with CIC-DARKNET2020 And make it public | 23 features | Characterize darknet traffic with 86% accuracy |
| Sridhar et al.[57] | 2021 | Flow, Packet | Semi-Supervised | RF, cGAN | CIC-DARKNET2020 | 20 features | Distinguished Tor traffic with an accuracy of 97.88% |
| Cai et al. [18] | 2019 | Flow, Packet | Semi-Supervised | (MMIRF & RF) for feature selection and XGB for classification | Anon17 | 18 features | Classified Tor, I2P and JonDonym with 99.73% accuracy |
| Zhao et al. [20] | 2021 | Flow | Deep Learning | CNN, LSTM | SJTU-AN21 ISCXVPN2016 and Anon17 | 35 features | Classify Tor, I2P and Jondonym traffic with an accuracy of 95.29% |
| Zhao et al. [44] | 2022 | Flow | Deep Learning | ResGCN | Anon17 | 50 features | Classify Tor, I2P and Jondonym traffic with an accuracy of 87.3% |
| Alimoradi et al. [9] | 2022 | Flow, Packet | Deep Learning | DNN | CIC-DARKNET2020 | 79 features | Detector classifies Tor-VPN traffic into 4 classes with an accuracy of 96% |
| Kumar et al. [36] | 2019 | Flow, Packet | Supervised | Microsoft Azure ML | Private data collected by Surfnet | 76 features | benign and malign traffic with 99% accuracy |
| Karunanayake et al. [6] | 2023 | Circuit, Flow, Packet | Supervised | KNN, RF, SVM | Public data of the Tor dataset consists of 95000 traffic traces the | 50 features | Detect Tor traffic from onion services with 99% accuracy |
| | | | | | OS dataset contains 41503. Also Generated WTFPAD, and TrafficSliver | | |
| Kim et al. [32] | 2018 | Packet | Deep Learning | CNN | UNB-CIC | 28 features | Classify Tor traffic with 99.3% accuracy |
| Vishnupriya et al.[80] | 2021 | Circuit, Flow, Packet | Deep Learning | RNN-LSTM | ISCXTor2016 | 28 features | Classify Tor traffic with 99.9% accuracy |
| Sarkar et al. [14] | 2020 | Circuit, Flow, Packet | Deep Learning | DNN | UNB-CIC | 25 features | Detected Tor traffic with the accuracy of 99.89% |
| Choorod et al. [31] | 2021 | Packet | DPI, Supervised Learning | J48, KNN, RF with 10 folds | CIC-DARKNET2020 | 16 features | Detected Tor and non-Tor based on a payload with 90% accuracy |
| Yin et al. [29] | 2022 | Flow, Packet | Deep Learning | CNN | Private data for training ISCXTor2016 for testing | Image features | Tor detection with an accuracy of 97.6% |

### B) PAYLOAD-BASED TECHNIQUES

The technique of analyzing the traffic that flows across a network and drawing conclusions about the people who use it is known as traffic analysis. The timing of packets as well as header information including source and destination IP addresses are examples of metrics utilized by these assumptions. Traffic analysis is a technique that can be applied within the context of anonymous networks to locate end-users striving to maintain their anonymity [81].

While a TCP packet is still inside the Tor network, traffic analysis cannot be used to read the header information or content of the original TCP packets that are being transported over the network. This is because the header information and content are encrypted before the packet enters the Tor network. Although it is possible to conduct this kind of traffic analysis both before and after the original packet enters and exits the Tor network, doing so is outside the purview of our threat model [48].

It is possible to conduct traffic analysis on the packets being transferred across the Tor network. An attacker will not be able to read the header information of the original TCP packets because they are encased in the onion routing packets; nonetheless, an attacker can still gather information based on the onion-routing packets themselves.

Cuzzocrea et al. [34] introduced a strategy for identifying Tor-related traffic originating on a host. Therefore, it can identify whether a user is utilizing the Tor application. The identification approach employs a supervised classification technique that relies on the characteristics of traffic flows. There are 23 selected attributes encompassing flow duration, flow bytes per second, flow inter-arrival time, and flow active time. The present investigation was conducted on a total of six machine algorithms, and it was found that the C4.5 technique yielded the highest level of accuracy among them.

Classification research has been conducted on the Anon17 dataset by Montieri et al. [19] employing four different classifier approaches: Naïve Bayes, Bayesian Network, C4.5, and Random Forest. The publicly available dataset encompasses traffic originating from three widely used anonymity services – Tor, I2P, and JonDonym. The writers conducted a three-tier classification process, starting with categorizing networks such as the Anon Network (including Tor, I2P, and JonDonym). This was followed by classifying traffic types, including Normal, Tor Apps, and I2P Apps. Lastly, the authors classified applications into categories such as Tor, Streaming, Torrent, and Browsing. The findings of this investigation indicate that it is possible to accurately identify and differentiate between all classification levels on anonymity services. The classification process involves utilizing 81 distinct features, which encompass various aspects such as flow direction, packet length, inter-arrival time, IP header features, and the count of connections seen over the lifespan of a given traffic flow. This experiment is distinguished from others due to its utilization of publically available datasets. However, given the dataset has only recently become public, there is currently a dearth of studies that use this particular dataset.

The DarknetSec method, which is a revolutionary self-attentive deep learning approach, has been presented for classifying darknet traffic and identifying applications [28]. The various components of DarknetSec are responsible for the analysis and handling of the payload content or payload statistics associated with a given network flow. In this study, a 1D CNN with self-attention embedding and a bidirectional Long Short-Term Memory (Bi-LSTM) network is utilized to extract local spatial-temporal features from the payload content of packets. A multi-head self-attention module is also developed to process the payload information simultaneously. The CICDarknet2020 dataset was utilized to apply the model, as it encompasses a comprehensive representation of darknet traffic, including both Virtual Private Network (VPN) and The Onion Router (Tor) applications. Comprehensive experimental results demonstrate the superiority of DarknetSec over other contemporary techniques, as evidenced by its remarkable multiclass accuracy of 92.22%. TABLE IX discussed some of the payload-based traffic analysis techniques.

| TABLE IX Summary of Payload-Based Traffic Analysis | | | | | | | |
|---|---|---|---|---|---|---|---|
| **Authors** | **Year** | **Traffic Attributes** | **Machine Learning Approaches** | **Technique** | **Data Type** | **Features** | **Output Features** |
| Jia et al. [78] | 2017 | Packet | Semi-Supervised | Decision Tree, Tri-Training algorithm | Private | 4 features | Accuracy is 94% for application categorization in Tor |
| Nhien et al. [82] | 2023 | Flow, Packet | Supervised | SVM, RF, GBDT, XGBoost, KNN, MLP,CNN and AG-CAN | CIC-DARKNET2020 | 61 features | 99.8% F1-score for traffic classification and 92.2% F1-score for application classification |
| Iliadis et al. [83] | 2021 | Flow, Packet | Supervised | KNN, MLP, RF, DT, GB | CIC-DARKNET2020 | 80 features | Detect Tor and VPN applications with an accuracy of 84.93% for MLP 98.21% for RF |
| Sarwar et al. [30] | 2021 | Flow | Semi-Supervised Deep Learning | DT, GB, RF, XGB, CNN-GRU, CNN-LSTM | CIC-DARKNET2020 | 20 features | Darknet traffic detection with an accuracy of 96% and 89% for darknet traffic application classification |
| Cai et al. [18] | 2019 | Flow, Packet | Semi-Supervised | (MMIRF & RF) for feature selection and XGB for classification | Anon17 | 18 features | Classified Tor, I2P and JonDonym with 99.73% accuracy |
| Yuzong et al. [62] | 2020 | Flow, Packet | Semi-Supervised Deep Learning | 6 ML algorithms LR, DT, RF, GBDT, XGBoost, LightGBM and 2 deep learning algorithms MLP, LSTM | Collected own data darknet-dataset-2020 and made it public | 26 features | Can detect Tor, I2P, Zeronet and Freenet with an accuracy of 96.9% and recognize 25 user applications with an accuracy of 91.6% |
| Shi et al. [67] | 2022 | Flow, Packet | Semi-Supervised | KNN | data darknet-dataset-2020 | 18 features | Can detect Freenet with an accuracy of 99.6%, 5 |
| | | | | | | | user behaviors with an accuracy of 95.8% |
| Shahbar et al. [22] | 2018 | Circuit, Flow | Supervised | RF, NB, C4.5 and Bayes network | ANON17 | 92 features | Classify Tor, I2P and Jondonym traffic and detected Tor pluggable transports with 94% accuracy |
| Sarkar et al. [14] | 2020 | Circuit, Flow, Packet | Deep Learning | DNN | UNB-CIC | 25 features | Detected Tor traffic with the accuracy of 99.89% |
| He et al. [17] | 2022 | Packet | Deep | RF, J48, IBK, BN, NB, MLP, CNN | ISCXTor2016 and PTO self-collected dataset AAT | MFD chromatographic features | Classify the interaction pattern between Tor applications with an accuracy of 88.3% |

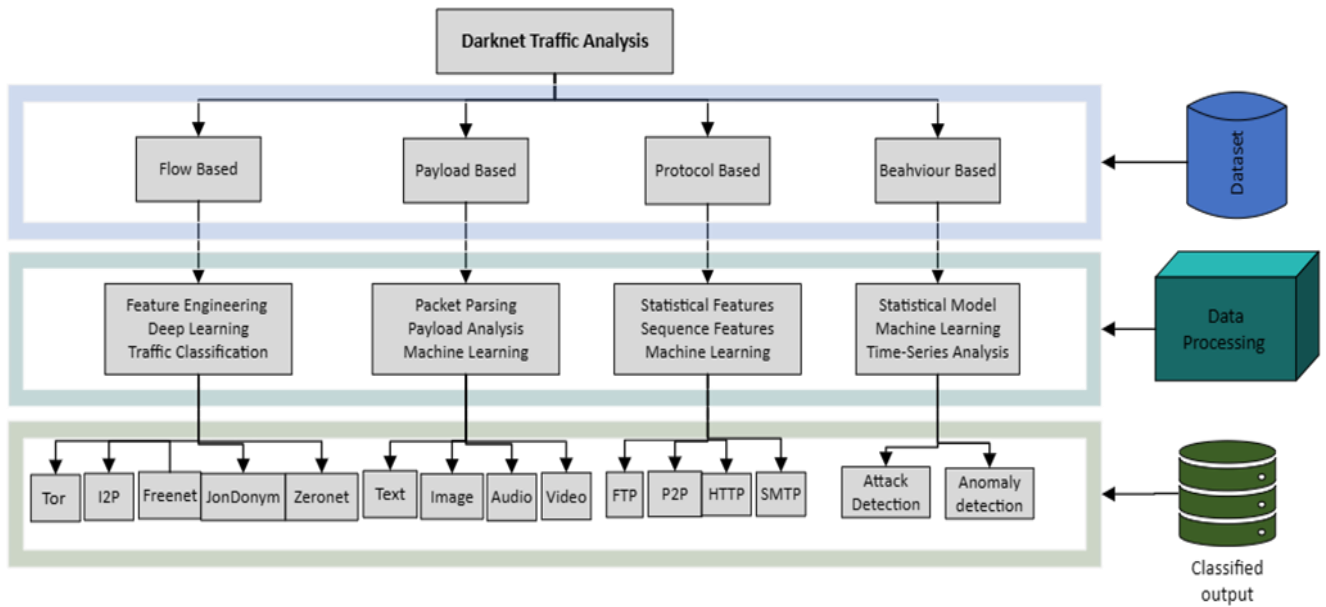

*Figure 13 Taxonomy of The Darknet Traffic Analysis*

### C) PROTOCOL-BASED TECHNIQUES

In their study, Guan et al. [15] employed the secure shell (SSH) protocol as the tunneling mechanism. They conducted an empirical investigation to assess the availability and traceability of Tor traffic that is tunneled through SSH. This research was carried out by employing machine learning algorithms that were trained using feature sets, focusing on traffic analysis. The feature sets provide the capability to depict the attributes of individual flows by considering two key aspects: payload pattern and statistical packet-level association. The initial phase of identification has two distinct components. The first component pertains to the many pluggable transports employed in tunneled Tor traffic. The second component relates to the diverse higher applications encompassed within a particular pluggable transport. Subsequently, the process of traffic correlation is executed on the inbound and outbound data streams that pertain to a certain tunneled Tor session, with the intention of compromising the level of anonymity provided by the system. Under specific situations, the accuracy and F1 scores exceed 95%, while the false positive rates tend to approach 0%.

### D) BEHAVIOR-BASED TECHNIQUES (RQ3.1)

The main goal of traffic analysis is to create algorithms and methods that can be used to observe, analyze, evaluate, and manage communication. In the case of anonymous networks like Tor, I2P, and Freenet, etc., traffic analysis is used to discover who is using the network and how they are using it. How well traffic analysis hacks work depends on how accurate the information the attacker has. The more network coverage an adversary model has, the more likely it is that the traffic being watched is correct. But when making an attack model, you should be aware of assumptions that aren't realistic about how long an observation will last and how much of the network will be covered. From the point of view of threat models, we will now talk about traffic analysis tactics. TABLE X shows the behavior-based traffic analysis.

### I. MALICIOUS TRAFFIC DETECTION

Dodia et al. [52] provided a comprehensive overview of traffic analysis methodologies that demonstrate high efficacy in precisely detecting and discerning communication patterns associated with Tor-based malware. The researchers gathered numerous Tor-based malware binaries, conducted execution and analysis on over 47,000 currently active encrypted malware connections, and subsequently compared these connections with benign browser traffic. In addition to conventional traffic analysis features, the authors additionally propose the incorporation of global host-level network features to effectively capture distinctive virus communication patterns within host logs. The trials provide confirmation that the models can identify previously unknown malware connections, commonly referred to as "zero-day" malware connections, with a false positive rate (FPR) of 0.7%. This ability remains consistent even when the proportion of malware connections within the Tor traces in the test set is less than 5%. By employing multi-labeling methodologies, they effectively identified several classes of malware based on their behavioral characteristics, including ransomware.

Kumar et al. [55] present a method for threat identification that involves monitoring dark web traffic through a machine learning classifier. The proposed system comprises four main components: traffic production and collection, feature extraction, dataset processing, and classifier design. The software SURFnet, renowned for its superior network capabilities, was employed for collecting darknet traffic, thereby ensuring the highest quality standard for research endeavors. The traffic data was collected by monitoring the traffic via the darknet sensor. Simultaneously, diverse applications such as Facebook, Twitter, and YouTube were employed to gather routine traffic. The approach involved the extraction of 76 characteristics from the generated traffic data, followed by using Microsoft Azure ML for pre-processing and training the machine learning model. The proposed system demonstrates the ability to accurately classify both malicious and benign traffic, achieving a precision rate of up to 99%.

### II. ADVERSAL ATTACK DETECTION

Adversarial attacks refer to introducing minor perturbations to training or inference samples. Adversarial attacks can manipulate darknet traffic data to render it indistinguishable from regular network data. The manipulation of darknet traffic data can evade detection by machine learning models. As a result, this can enable attackers to compromise network nodes and engage in illicit activities, such as causing economic losses, terminating operations, and leaking confidential information [26]. In recent times, there has been a notable utilization of machine learning and deep learning (ML/DL) methodologies in several safety-critical domains, including network security systems. Two notable instances within the realm of network security encompass intrusion detection systems (IDS) and darknet traffic classification systems. Nevertheless, new research suggests that machine learning and deep learning systems are susceptible to minor adversarial perturbations introduced to the input data. These attacks are commonly referred to as adversarial attacks in academic literature. The potential alterations can significantly impact the effectiveness of the machine learning and deep learning models employed in network security solutions [26].

In recent times, there has been a surge in the utilization of Generative Adversarial Networks (GAN) for generating adversarial samples to enhance the security of network systems. Usama et al. [84] have presented a novel attack and defense technique utilizing Generative Adversarial Networks (GANs). The researchers employed a GAN based an adversarial attack technique to infiltrate an intrusion detection system (IDS) while minimizing alterations to the network traffic characteristics. The authors modified the defense mechanism based on Adversarial Training (AT) by incorporating Generative Adversarial Networks into the model pipeline. Furthermore, the model underwent training using a combination of adversarial samples provided by the defender and the Generative Adversarial Network (GAN), encompassing known and unanticipated types. The employed technique enhanced the model's resilience against various forms of adversarial attacks employed by malicious actors to deceive the model.

Mohanty et al. [26] presented a Stacking Ensemble (SE) model that aims to optimize the integration of predictions from three base learners, namely Random Forest (RF), K-Nearest Neighbors (KNN), and Decision Tree (DT), to enhance the overall performance of darknet characterization. The proposed approach involves the development of a two-layered autoencoder-based defense mechanism, which consists of a detector and denoizer. This mechanism aims to enhance the resilience of the network security system against adversarial attacks. The efficacy of the proposed approach is showcased through a comprehensive series of experiments conducted on the CIC-Darknet-2020 dataset. The model's resilience is evaluated by subjecting it to three widely applicable adversarial attacks, namely the Fast Gradient Sign Method (FGSM), the Basic Iterative Method (BIM), and DeepFool. Additionally, a realistic Boundary assault is also employed for testing. The experimental findings indicate that the SE model performs better than the baseline Deep Image and other competing models. Specifically, it achieves an accuracy score of 98.89% in the context of identifying darknet traffic [26].

| TABLE X Summary of Behavior-Based Traffic Analysis ||||||||
|---|---|---|---|---|---|---|---|
| **Authors** | **Year** | **Traffic Attributes** | **Machine Learning Approaches** | **Technique** | **Data Type** | **Features** | **Output Features** |
| Kumar et al. [36] | 2019 | Flow, Packet | Supervised | Microsoft Azure ML | Private data collected by Surfnet | 76 features | Benign and malign traffic with 99% accuracy |
| Dodia et al. [85][45] | 2022 | circuit | Semi-Supervised | LGBM, XGB, RF, KNN, LR, Extra Trees, CatBoost | Used VirusTotal to source malware corpus | Used top 10 features, 9 host level and one connect level feature | Tor-based malware detection with 0.7% FPR |
| Mohanty et al. [26] | 2022 | Flow, Packet | Semi-Supervised | Stacking Ensemble Model (RF, KNN, DT) along with two stage autoencoders | CICDarknet-2020 | 79 features | Presented application level classification under the traffic adversarial attack settings |
| Ban et al. [86] | 2017 | Packet | Semi-Supervised | SVM, AE | Data gathered through NICTER | -- | Detected around 3000 botnet occurrences on Port 80 and DDos event with 96.7% accuracy |

### III. DETECTION OF BOTNET

A botnet refers to a collection of hacked hosts under the remote control of botmasters, who utilize them for malicious purposes, including launching denial of service attacks, stealing personal information, and engaging in spamming operations. Botmasters exert control over their botnets by utilizing a network of Command and Control (C&C) servers. In the context of bot management, it is observed that a botmaster, who typically oversees a substantial number of bots, experiences a consistent pattern wherein the bots exhibit a synchronized reaction to commands issued by the botmaster. If the darknet successfully captures the probe, it is expected that many hosts exhibiting comparable behaviors will manifest in a synchronized manner. This hypothesis provides a plausible rationale for the heightened occurrence of coordinated probing activities seen by clusters of hosts. Examining the coordinates could potentially unveil the underlying factors contributing to the probes, hence enabling the calculation of the population of probing hosts [45].

The actions of botnets exhibit a significant temporal correlation that is readily detectable within the darknet. The utilization of abrupt change detection can be employed to extract botnet events with a high rate of detection. Ban et al. [45] proposed an active Epoch detection using a Cumulated Sum algorithm to detect botnet activities. The temporal characteristics, specifically the degree of coincidence across hosts, include significant discriminative information that can be utilized to identify hosts involved in coordinated botnet probes.

### IV. DETECTION OF DDOS ATTACKS

By studying backscatter packets obtained from the darknet as a response to a DDoS attack that occurred on the internet [45], it is possible to accurately detect Distributed Denial of Service (DDoS) occurrences. The suggested system does feature extraction on a probing host by utilizing packets received from the host and afterwards employs supervised learning techniques to forecast the host's status. DDoS incidents can be detected with a rather high level of accuracy, specifically 96.7%, even without implementing incremental learning. All instances of Distributed Denial of Service (DDoS) attacks are accurately identified.

V. CORRELATION ATTACKS

Correlation attacks are specifically devised to identify and analyze the patterns and associations in communication interactions between clients and servers. End-to-end attacks can manifest in either an active or passive form. In the context of these assaults, the adversary engages in the surveillance of entrance and exit nodes situated at both ends [61].

Tor cells follow a specific sequence as they pass through various nodes, including the source, entry guard, middle relay, exit relay, and finally the destination. The phenomenon of traffic pattern alterations, when observed at one end of a given path, will inevitably manifest at the opposite end of that path. Consequently, opponents who are present at both ends of these paths possess a significant likelihood of successfully deanonymizing users. The asymmetric characteristic of internet routing exacerbates the vulnerability of Tor traffic correlation assaults by augmenting the likelihood of intercepting user traffic at both ends, at least in one direction [50]. The authors introduced a Tor path selection method that considers the distance between nodes to address traffic correlation threats. The objective of the technique is to reduce the likelihood of traffic correlation attacks without necessitating any prior knowledge of AS-level route patterns. The datasets are generated by CollecTor, which can regenerate a Tor network state at a specific time. The findings from the simulation indicate that the distance-aware algorithm offers a significant improvement in mitigating the risk of Tor traffic correlation assaults. Specifically, it achieves a reduction of up to 27% compared to the currently employed AS-aware method. Moreover, the distance-aware strategy outperforms the AS-aware algorithm in around 88% of the evaluated scenarios.

E. COUNTERMEASURES (RQ3.2)

To mitigate deanonymizing attacks, several countermeasures have been implemented to effectively respond to or enhance resistance against such attacks.

a. PACKET PADDING

Packet padding is commonly employed to augment the size of packets to exclude the impact of packet attributes, such as packet order and packet size, from descriptions. By employing this methodology, it becomes feasible to effortlessly adjust the size of each packet to match the precise dimensions, such as the maximum transmission unit (MTU). Similarly, multiple methodologies have been examined in diverse research endeavors to optimize the padding of packet size with both efficiency and efficacy [61]. To enhance the efficiency of traffic flow, artificial delays might be introduced between each packet to obscure the measurement of traffic time. A number of padding schemes are discussed below.
- Dummy padding is an approach that involves the deliberate insertion of dummy packets into the original traffic of users to obscure the actual volume of traffic.
- Delayed padding is a relay-based padding strategy that serves as an alternate approach. The aforementioned system employs a hybrid approach involving the manipulation of packet delays and the insertion of dummy packets to obscure the temporal patterns present in packet flows. Delays are managed by implementing a stringent upper limit to restrict the duration of the delays applied. Likewise, including dummy packets is facilitated by employing a minimum transmission rate [42].

**TABLE XI Taxonomy Of Darknet Traffic Analysis**

| Ref. ID | Flow-Based | | | | | Behavior-Based | Protocol-Based | Payload-Based |
|---|---|---|---|---|---|---|---|---|
| | Tor | I2P | Freenet | JonDonym | Zeronet | Attack detection | Protocol detection | App detection |
| ID[9] | √ | | | | | | | |
| ID[57] | √ | | | | | | | |
| ID[67] | | | √ | | | | | |
| ID[18] | √ | √ | | √ | | | | |

| ID | | | | | | | |
|---|---|---|---|---|---|---|---|
| ID[12] | √ | | | | | | |
| ID[87] | √ | √ | | √ | | | |
| ID[19] | √ | √ | | √ | | | √ |
| ID[85] | | | | | √ | | |
| ID[5] | | | | | √ | | |
| ID[10] | √ | | | | | | |
| ID[82] | √ | | | | | | |
| ID[75] | | | | | √ | | |
| ID[88] | √ | | | | | | |
| ID[79] | √ | | | | | | |
| ID[55] | | | | | √ | | |
| ID[8] | √ | | | | | | |
| ID[38] | √ | | | | | | |
| ID[32] | √ | | | | | | |
| ID[69] | √ | | | | | | |
| ID[89] | √ | | | | | | |
| ID[85] | √ | | | | √ | | |
| ID[39] | | | | | √ | | |
| ID[15] | √ | | | | | √ | |
| ID[53] | | | | | √ | | |
| ID[30] | √ | | | | | | |
| ID[90] | √ | | | | | | |
| ID[91] | | | | | | | √ |
| ID[70] | √ | | | | | | |
| ID[92] | √ | | | | | | |
| ID[13] | √ | | | | | | |
| ID[56] | √ | | | | | | |
| ID[28] | √ | | | | | | √ |
| ID[83] | √ | | | | | | |
| ID[15] | √ | | | √ | | | |
| ID[23] | √ | √ | √ | √ | | | |
| ID[48] | √ | | | | | | |
| ID[26] | √ | | | | √ | | |
| ID[80] | √ | | | | | | |
| ID[22] | √ | √ | | √ | | | |
| ID[93] | √ | √ | | √ | | | |
| ID[14] | √ | | | | | | |
| ID[68] | √ | | | | | | |
| ID[31] | √ | | | | | | |
| ID[33] | √ | | | | | | |
| ID[29] | √ | | | | | | √ |
| ID[17] | √ | | | | | | |

### c. TRAFFIC MORPHING

Traffic morphing is a technique that can be utilized to alter the appearance of traffic to deviate from its inherent pattern. To counteract a website fingerprinting assault, a potential strategy for the web server involves the initial selection of a certain page to serve as the target [61]. Subsequently, the server can imitate the distribution of packet sizes as a means of defense. The transmission of packets occurs uniformly throughout the network, effectively eliminating any discernible temporal patterns. In addition, the technique incorporates packet dropping to obscure the perceived quantity of transmitted packets. Specifically, dummy packets are deliberately chosen to be discarded by particular relays across the circuit [42].

### d. BANDWIDTH VERIFICATION

Better verification of bandwidth, bandwidth authorities (bwauths) make sure that malicious relay operators don't make false claims about their bandwidth and change the method for choosing the relay or path. Basically, the "bwauths" actively probe relays at regular intervals, recording their observed bandwidth capabilities, comparing them to similar-level relays, and then adjusting the weights accordingly. This causes the bandwidth weight of a relay to slowly rise over time until it peaks and then falls [42].

## VII. CHALLENGES & LIMITATIONS IN THE DARKNET TRAFFIC ANALYSIS (RQ4)

The literature offers valuable insights into the importance of precise and effective identification and analysis of darknet traffic. However, there remain specific areas that require further attention and investigation. One of the main obstacles encountered in the analysis of darknet traffic relates to the encryption of packets, hence achieving the extraction of significant insights from the traffic data is a challenging task. The existing detection methods have a restricted scope when it comes to handling encrypted network traffic, due to the ongoing growth of encryption algorithms employed in darknet traffic, hence posing difficulties in developing efficient detection and classification models.

Another obstacle is the availability of quality datasets employed for training and assessing algorithms designed to analyze darknet traffic. The lack of sufficient access to empirical darknet traffic data poses a significant constraint on the capacity to construct reliable and widely applicable models. In addition, the current detection methods exhibit limitations in their ability to analyze a limited range of darknet data, including tunnel network traffic and anonymous network traffic. Consequently, these technologies are inadequate in efficiently detecting and categorizing newly emerging forms of darknet traffic.

Furthermore, the elevated computational difficulty of deep learning models utilized in analyzing darknet traffic presents a formidable obstacle in relation to scalability and the ability to handle data in real-time. To address these issues, it is recommended that future research focuses on the development of flexible and scalable deep learning models specifically designed for the analysis of darknet data.

The traffic analysis models are required to have the ability to consistently adapt to emerging encryption schemes and evolving forms of darknet traffic. In addition, it is imperative to make new comprehensive datasets publically available for researchers to train and test models with improved efficacy. In addition, it is important to investigate novel methodologies to mitigate the computational intricacy associated with deep learning models, aiming to facilitate effective real-time processing of darknet traffic.

In summary, the domain of darknet traffic analysis encounters many kinds of obstacles that necessitate solutions. The primary obstacles encountered in the study of darknet traffic are the persistent advancement of encryption methodologies, the restricted availability of empirical data, and the application of existing models in all types of anonymous traffic to accurately identify and categorize darknet traffic.

## VIII. RESEARCH GAP AND FUTURE WORK (RQ5)

In addition to the study directions highlighted within this work, various domains are worth exploring for prospective research in the realm of darknet traffic analysis. Firstly, there is potential for conducting additional research on integrating machine learning algorithms with deep learning techniques. The integration of deep learning techniques with conventional machine learning algorithms could improve the effectiveness and precision of detection models for darknet traffic.

This might include investigating ensemble approaches or hybrid models that capitalize on the advantages of deep learning and classical machine learning techniques. Additionally, it is imperative to acknowledge the challenge posed by the restricted accessibility of empirical data from the physical world. Researchers must prioritize the acquisition and management of a wide range of contemporary datasets that effectively capture the intricate nature of darknet activity in the actual world. This may entail collaborating with law enforcement agencies, network administrators, and other professionals to acquire access to pertinent data for analysis. Furthermore, it is imperative to establish robust evaluation measures for studying darknet traffic. The measurements should consider the distinctive attributes and complexities associated with darknet traffic, encompassing unbalanced data and the continuous evolution of attack strategies. Furthermore, it is imperative to investigate the implementation of privacy-preserving methodologies in analyzing darknet traffic. The protection of sensitive information while enabling efficient analysis and detection of darknet traffic can be achieved through techniques such as differential privacy or homomorphic encryption. In addition, future research must emphasize the advancement of real-time detection and analysis methodologies for darknet traffic.

The proposed approaches entail the implementation of streaming data processing frameworks and the optimization of algorithms to effectively manage the substantial volume and rapid flow of darknet traffic. In conclusion, further investigation in darknet traffic analysis should delve into the combination of machine learning algorithms and deep learning techniques, address the challenge of the availability of comprehensive real-world data, employ privacy-preserving methods, and prioritize the examination and assessment of real-time detection and analysis. In conclusion, it is recommended that forthcoming investigations in the domain of darknet traffic analysis prioritize the integration of machine learning algorithms with deep learning approaches to augment the efficacy of detection models.

Choosing the right features is an important part of analyzing darknet data. The huge number of possible features and the fact that darknet traffic always changes make choosing features difficult and often confusing. Researchers have difficulty figuring out which traits are most important for correctly classifying traffic, which results in less-than-ideal analysis results and limited detection abilities. More work is needed to design efficient feature selection algorithms that lead to more efficient classification and quicker model processing.

Furthermore, it is advisable to focus more on examining encrypted darknet communication. With the increasing sophistication of encryption technologies, designing innovative approaches that can proficiently examine and comprehend the substance of encrypted data flow is imperative.

## IX. CONCLUSION

This systematic literature review offers a thorough exposition of the darknet's traffic, encompassing its technical and forensic complexities related to anonymous network topologies. Additionally, it explores the various approaches, algorithms, tools, and strategies employed for detecting and identifying darknet traffic.

The survey examines the strategies and methodologies employed in these works and their inherent constraints. The subject matter encompasses a range of network traffic analysis facets, such as categorizing protocols and applications, identifying application usage patterns, and examining quality and user experience within encrypted networks. The traffic analysis approaches have been presented in Tables VIII, IX, and X. This paper presents a comprehensive study and simplification of numerous attacks and countermeasure approaches. The purpose of this text is to present a comprehensive overview of the current literature in this discipline, highlighting any gaps, limitations, and opportunities for further advancement.

The topic at hand involves various challenges. Accurately detecting and classifying developing types of darknet traffic presents substantial challenges due to the ongoing advancement of encryption techniques and the restricted availability of real-world data.

To address these constraints, researchers have utilized sophisticated methodologies, such as deep neural networks and self-attention mechanisms. These methodologies have demonstrated potential in the automated extraction of advanced information from network data, enhancing the precision of classification and detection.

The rapid evolution of the darknet traffic monitoring sector necessitates the use of adaptive and dynamic methodologies to effectively respond to the ever-changing environment of darknet activity. In addition, it is imperative to emphasize the significance of fostering collaboration among scholars, industry professionals, and law enforcement entities to acquire authentic darknet traffic statistics, exchange specialized knowledge, and devise efficacious strategies. This comprehensive literature review provides an overview of the current issues associated with analyzing darknet traffic and proposes potential future directions for research in this domain.

| List of Acronyms | |
|---|---|
| Artificial Neural Network (ANN) | ANN |
| Auxiliary-Classifier Generative Adversarial Networks | AG-CAN |
| Conditional Generative Adversarial Network | cGAN |
| Convolutional Neural Networks | CNN |
| Convolution-Gradient Recurrent Unit | CNN-GRU |
| Convolution-Long Short-Term Memory | CNN-LSTM |
| Decision Tree | DT |
| Deep Brief Networks | DBN |
| Deep Neural Network | DNN |
| Extreme Gradient Boosting | XGB |
| Gradient Boosting | GB |
| Gradient Boosting Decision Trees | GBDT |
| K-Nearest | KNN |
| Long Short-Term Memory | LSTM |
| Mount Frequency Direction | MFD |
| Multidimensional Scaling | MDS |
| Multilayer Perceptron | MLP |
| Partial Latest Square Regression | PLS |
| Principal Component Analysis | PCA |
| Random Forest | RF |
| Random Forest | RF |
| Residual Graph Convolutional Networks | ResGCN |
| Support Vector Machine | SVM |